\documentclass[twocolumn]{aastex631}

\begin{document}

\title{The Dog that Didn't Bark: Non-Variable Field Stars in the RR Lyrae Instability Strip}

\correspondingauthor{Yuxi(Lucy) Lu}
\email{lucylulu12311@gmail.com}

\newcommand{\osu}{Department of Astronomy, The Ohio State University, Columbus, 140 W 18th Ave, OH 43210, USA}
\newcommand{\ccapp}{Center for Cosmology and Astroparticle Physics (CCAPP), The Ohio State University, 191 W. Woodruff Ave., Columbus, OH 43210, USA}
\newcommand{\bpg}{$(G_{\rm BP}-G)$}
\newcommand{\mg}{$M_G$}
\newcommand{\bpgo}{$(G_{\rm BP}-G)_0$}
\newcommand{\mgo}{$(M_G)_0$}

\author[0000-0003-4769-3273]{Yuxi(Lucy) Lu}
\affiliation{\osu}
\affiliation{\ccapp}

\author[0000-0002-6330-2394]{Cecilia Mateu}
\affiliation{Departamento de Astronomía, Facultad de Ciencias, Universidad de la República, Iguá 4225, 14000 Montevideo, Uruguay}

\author[0009-0001-1470-8400]{K. Z. Stanek}
\affiliation{\osu}
\affiliation{\ccapp}

\begin{abstract}
RR Lyrae stars (RRLs) are easy to identify thanks to their large photometric variation and short periods.
All stars in the RRL instability strip are pulsators is often a hidden assumption in most stellar population studies using RRLs.
Non-variable stars in the instability strip have been discovered for Cepheids and $\delta$ Scuti, and in this paper, we report the discovery of non-variable filed stars in the RRL instability strip.
Using a high-quality sample selected from Gaia DR3, we find at least 15\% of the stars in the empirical instability strip where the variable fraction is $>$ 0.7 have near-zero photometric variations or variations that are significantly smaller than typical RRLs. 
The non-variable stars are mostly bright and close by, on cold orbits in the Galactic plane.
Metallicity from Gaia BP/RP spectra suggests the non-variable stars have an average metallicity is $\sim$ -0.5 dex, with a peak at 0. 
The discovery of these non-variable stars in the RRL instability strip challenges our understanding of stellar physics and further investigation is needed to understand the origin of these stars. 
\end{abstract}



\section{Introduction} \label{sec:intro}
RR Lyrae stars (RRLs) are helium-burning horizontal branch stars that sit in the instability strip.
RRLs occupy a narrow range of absolute mean magnitudes and exhibit a period--luminosity--metallicity relation, making them useful for distance measurements \citep[e.g.,][]{Caputo2000, Bono2003, Garofalo2022}.
RR Lyrae instability is driven by He+ ionization \citep[e.g.,][]{Cox1963, Stellingwerf1982, Marconi2015}, and can be mostly classified as fundamental-mode pulsators (RRab) and first-overtone pulsators (RRc), with periods in the range between about 0.2 day and 0.8 day.
Another category of RRLs, RRd, which have two simultaneously excited pulsation modes, are relatively rare but also exist. 
RRab stars are easy to identify as they are bright, and have large photometric variations with distinct shapes.
However, due to lower amplitude and the near sinusoidal light curve shapes, RRc stars are often misclassified as other variables, usually eclipsing contact binaries.

That all stars in the instability strip should pulsate is oftentimes a hidden assumption that goes into pulsation models, not just of RRLs but of all kinds of pulsators within the strip. On his comprehensive book on RRLs, \citet{Smith1995} notes this notion dates back to Schwarzschild’s early studies about stellar pulsation:  ``Surmizing that something about this zone not only permitted the existence of RRLs but made it mandatory that stars within the zone pulsate, Schwarzschild concluded that `a star which can pulsate does pulsate’ ’’.  Understanding whether this conjecture holds has implications for correctly modeling the makeup of stellar populations and their variable star content, and therefore correctly predicting their observed properties via stellar populations via synthesis models, both for resolved and unresolved populations.

Non-variable field stars in the instability strip have been discovered for the Cepheids \citep{Andrievsky1996, Narloch2019} and $\delta$ Scuti \citep[e.g.,][]{Murphy2015, Murphy2019}.
Surprisingly, there are only a limited number of studies that mention non-variable stars in the RRL instability strip.
\cite{Cox1973} shows Population II RRLs and Cepheids with helium content, $Y$, less than 0.2 will not vary.
They concluded $Y$ must be greater than 0.22 for Population II stars as no non-variable RRLs were observed. 
However, \cite{CruzReyes2024} examined stars in the instability strip along the horizontal branch in globular clusters and found that 25\% of the stars located in the observational instability strip do not exhibit variability.

With large-scale photometric surveys such as Gaia \citep{gaia}, Kepler \citep{kepler}, TESS \citep{TESS}, ASAS-SN \citep{asassn}, and ZTF \citep{ZTF}, astronomers were able to construct large catalogs of field RRLs \citep[e.g.,][]{Jayasinghe2019, Huang2022, Molnar2022, Clementini2023}.
These catalogs can be used to revisit the question of whether non-variable field stars exist in the RRL instability strip.
In this paper, we answer this question by merging astrometric information from Gaia DR3 \citet{gaiadr3} with the three largest public RRL catalogs: Gaia SOS, ASAS-SN, and PS1 \citep{Clementini2023,Jayasinghe2019,Sesar2017_ps1_rrl}.
We find $\sim$ 30\% of the field stars in the center of the observational instability strip are not varying. 
The data selection is described in Section~\ref{sec:datamethod}, and exploratory investigations on the possibility of contamination, missed detection, and physical origin are done in Section~\ref{sec:results}.

\section{Data \& Selection} \label{sec:datamethod}
\subsection{The RR Lyrae catalog}\label{subsec:rrlcat}
We use a catalog comprising >300K stars that combines the three largest public surveys of RRL stars currently available: Gaia DR3 Specific Objects Study \citep{Clementini2023},  PanSTARRS-1 \citep{Sesar2017_ps1_rrl} and ASAS-SN \citep{Jayasinghe2019}. As described in \citet{Mateu2024}, the combination of these three catalogs provides the best performance in terms of sky coverage, depth, and completeness, empirically determined by probabilistic arguments from the three catalogs themselves following \citet{Mateu2020}. Details on how the three surveys were combined can be found in \citet{Mateu2024} and \citet{CabreraGadea2024b}.

\subsection{Data selection for Gaia DR3} \label{subsec:gaiadata}
To find whether non-variable stars in the RRL instability strip exist, we first obtained a high-quality sample of stars in and near the RRL instability strip from Gaia DR3 \citep{gaiadr3}.
Since we want to include stars that are undoubtedly in the empirical instability strip, we selected stars with low extinction and good parallax measurements.  
The quality cuts we did to obtain our sample from the full Gaia DR3 dataset and their purpose are shown below\footnote{For a list of Gaia DR3 column descriptions, see \url{https://gea.esac.esa.int/archive/documentation/GDR3/Gaia_archive/chap_datamodel/sec_dm_main_source_catalogue/ssec_dm_gaia_source.html}.}:
\begin{itemize} 
    \item Quality cut to ensure low extinction (cuts on \texttt{ag\_gspphot}), good parallax (cuts on \texttt{parallax\_over\_err}), and accurate photometric solutions (cuts on \texttt{ruwe}).
    \begin{itemize}
        \item \texttt{ag\_gspphot} $<$ 0.2
        \item \texttt{parallax\_over\_err} $>$ 20
        \item \texttt{ruwe} $<$ 1.4
    \end{itemize}

    \item The completeness of RRc is low as they have relatively low amplitudes and can be easily misclassified as eclipsing binaries \citep{Mateu2020}, this is especially true at low Galactic latitudes where crowding is significant. To ensure completeness for all RRLs and avoid possible extinction map failure modes in the Galactic plane, we also performed cuts on Galactic latitude:
    
    \begin{itemize}
        \item $|$\texttt{b}$|$ $>$ 5 deg
    \end{itemize}

    \item Photometric quality flags suggestions by \cite{gaia2018HR}
    \begin{itemize}
    \item \texttt{phot\_BP\_RP\_excess\_factor} $<$ \\ 1.3+0.06*power(\texttt{phot\_BP\_mean\_mag}-\texttt{phot\_RP\_mean\_mag},2)
    \item \texttt{phot\_BP\_RP\_excess\_factor} $>$ \\ 1.0+0.015*power(\texttt{phot\_BP\_mean\_mag}-\texttt{phot\_RP\_mean\_mag},2)
    \item \texttt{phot\_G\_mean\_flux\_over\_err} $<$ 50
    \item \texttt{phot\_bp\_mean\_flux\_over\_err} $<$ 20
    \item \texttt{phot\_rp\_mean\_flux\_over\_err} $<$ 20
    \end{itemize}

    \item Selecting the area of the Color--Magnitude--Diagram (CMD) close to the empirical RRLs instability strip.
    \begin{itemize}
    \item -0.5 $<$ \mg\ $<$ 1.5
    \item 0 $<$ \bpg\ $<$ 0.4
    \end{itemize}
\end{itemize}
In which \bpg\ is the Gaia color calculated by subtracting the $G$--band magnitude from the BP--band magnitude, and \mg\ stands for the absolute G magnitude and is calculated as \mg\ = \texttt{phot\_G\_mean\_mag} - 5*$\log_{10}({1/\texttt{parallax}})$-10.
These quality cuts left us with 19,148 stars, shown as the background histogram in Figure~\ref{fig:fig1} top plot.
It is worth noting that most of these stars do not lay in the RRL instability strip, thus, explaining the overly large number of stars in this sample. 

\subsection{Non-variable and variable RR Lyrae stars in the instability strip} 
\begin{figure}
    \centering
    \includegraphics[width=\columnwidth]{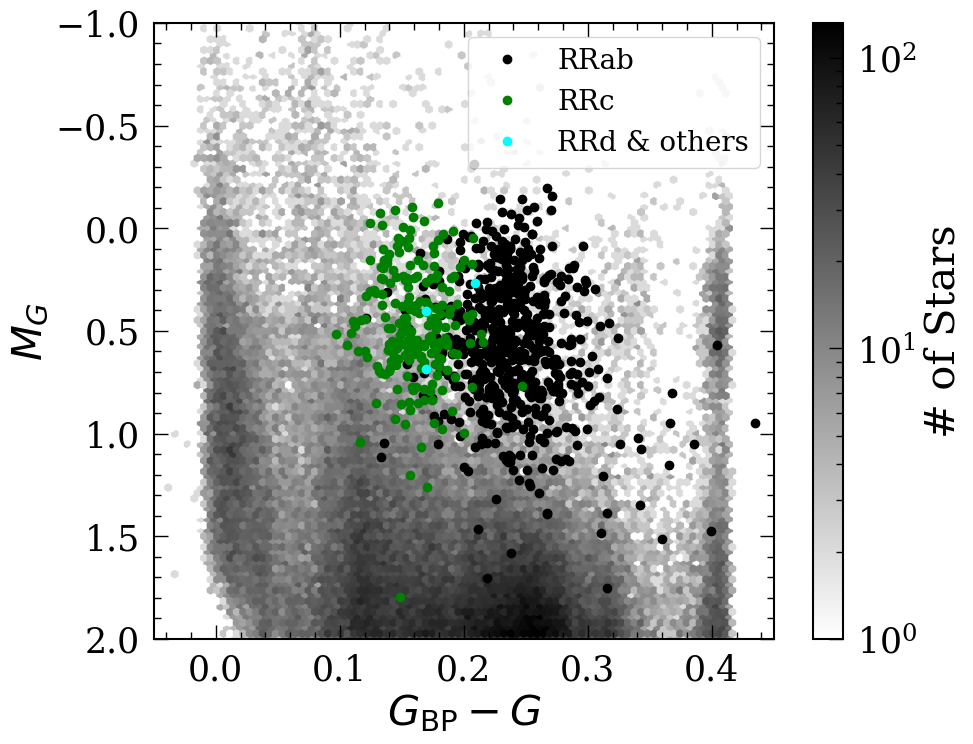}
    \caption{The background grey histogram shows the CMD of the 19,148 Gaia DR3 stars around the RRL instability strip with rigorous selection on extinction and parallax error. The points show the 750 literature-identified RRLs within the Gaia DR3 sample, colored by their subcategories.}
    \label{fig:fig1}
\end{figure}

After selecting our low extinction, low parallax uncertainty stars from Gaia DR3, we then cross-matched this sample to the RRLs catalog described in Section~\ref{subsec:rrlcat} using a cone search with 1.5'' search radius, which provided us with 750 RRLs. 
These RRLs are shown as points in Figure~\ref{fig:fig1} top plot, colored by their subcategories.

Even though extinction from interstellar dust shouldn't significantly affect the stars' positions in the CMD for our sample given the rigorous selection criteria, we still corrected for extinction using the \texttt{Bayestar19} 3D dust map \citep{Green2019} implemented in the \texttt{dustmaps} Python package \citep{Green2018}.
The extinction coefficients are taken from \cite{Danielski2018}\footnote{https://www.cosmos.esa.int/web/gaia/edr3-extinction-law}.
The average extinct values in \bpg\ and \mg\ for our sample of stars are close to 0, and the maximum extinction values in \bpg\ and \mg\ are 0.005 mag and 0.022 mag, respectively. 
The maximum extinction values are shown as the red error bar in the left plot of Figure~\ref{fig:fig2}.
Figure~\ref{fig:fig1} but plotted in dereddened \bpg\ (from now on noted by \bpgo) and dereddened \mg\ (from now on noted by \mgo) is shown in Figure~\ref{fig:fig2} left plot. 

To calculate the variable fraction, we first divided \bpgo\ and \mgo\ into 20 bins with equal spacing, with \bpgo\ ranging from 0 to 0.4 mag, and \mgo\ ranging from -0.5 mag to 1.5 mag. 
This gave us a bin width of 0.02 mag in \bpgo\ and 0.11 mag in \mgo, which both are significantly larger than the maximum extinction values for our sample. 
The variable fraction is then calculated for each bin by dividing the total number of RRLs in that bin by the total number of stars in the same color and magnitude range. 
The variable fraction for RRLs as a position on the CMD is shown in the right plot of Figure~\ref{fig:fig2}.
Only bins with more than 5 stars are plotted.

\section{The Dog that Didn't Bark} \label{sec:results}
The calculations we did in the last section suggest that the very center of the empirical RRL instability strip with a width of $\sim$ 0.005 mag in \bpgo\ contains only RRLs.
However, the variability fraction decreases dramatically and $\sim$15\% of the stars in bins with variable fraction $>$ 0.7 are non-variable.
The non-variable fraction increases to $\sim$30\% for stars in bins with variable fraction $>$ 0.5.
This fraction agrees with what was found in \cite{CruzReyes2024} for RRLs in globular clusters. 
In this section, we investigate the possibility of systematic biases (Section~\ref{subsec:contamination}, \ref{subsec:tess}) and physical origin (Section~\ref{subsec:phy}) of these non-variable stars in the RRL instability strip.

\begin{figure*}
    \centering
    \includegraphics[width=0.46\linewidth]{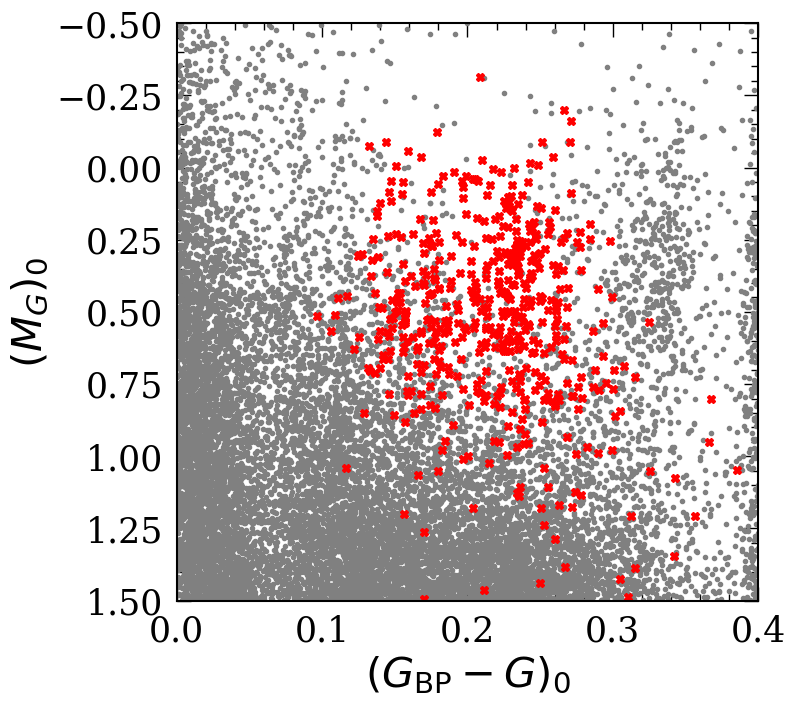}
    \includegraphics[width=0.53\linewidth]{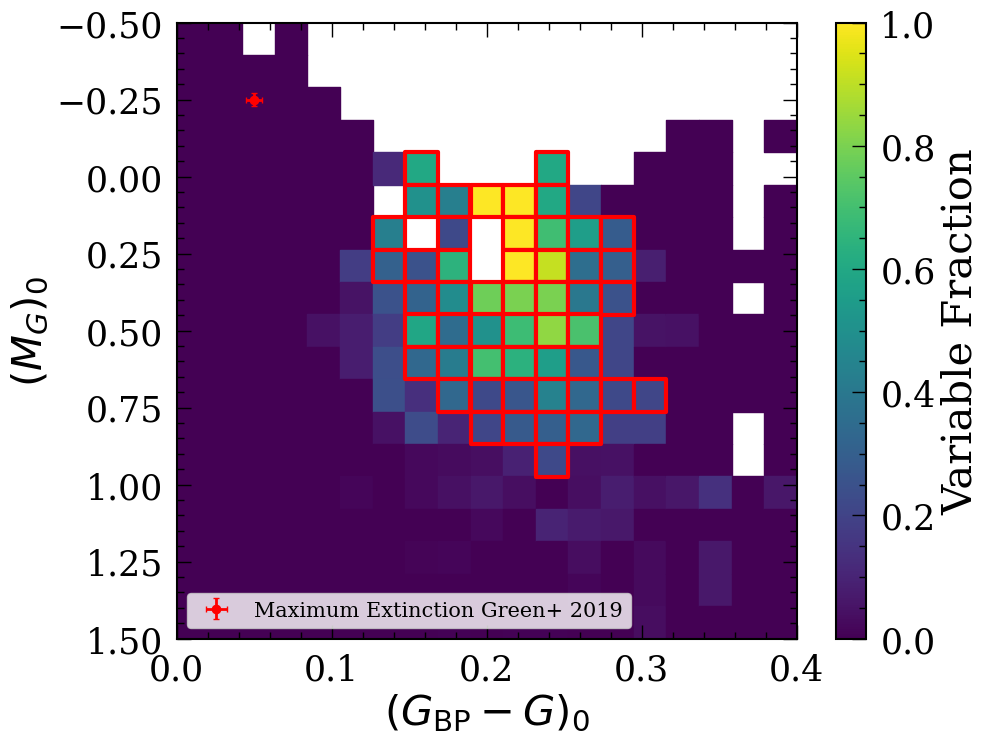}
    \caption{Left: \bpg\ and \mg\ after corrected for extinction using the \texttt{Bayestar19} 3D dust map \citep{Green2019} implemented in the \texttt{dustmaps} Python package \citep{Green2018} for the full Gaia DR3 sample (grey points) and the literature RRLs (red points).
    Right: a histogram showing the variable fraction as a position on the CMD. 
    The red error bar on the top left corner shows the maximum extinction values (0.005 mag in \bpg\ and 0.022 mag in \mg) for stars in the full Gaia DR3 sample.
    The red outlined bins show the ones with variable fraction $>$ 0.2.}
    \label{fig:fig2}
\end{figure*}

In the rest of this section, unless stated otherwise, the non-variable stars are defined to be those that are not in the RRLs catalog but occupy the \bpgo--\mgo\ bins with $>$0.2 variable fraction (shown as the red outlined bins in Figure~\ref{fig:fig2} left plot), and the variable stars are defined to be the RRLs that are in the same CMD bins.
This gave us 416 non-variable stars in the RRL instability strip. 
We also performed the same tests with stars in \bpgo--\mgo\ bins with $>$0.5 variable fraction and found no significant change to our results.

\subsection{Contamination from other field stars?}\label{subsec:contamination}
\begin{figure}
    \centering
    \includegraphics[width=\columnwidth]{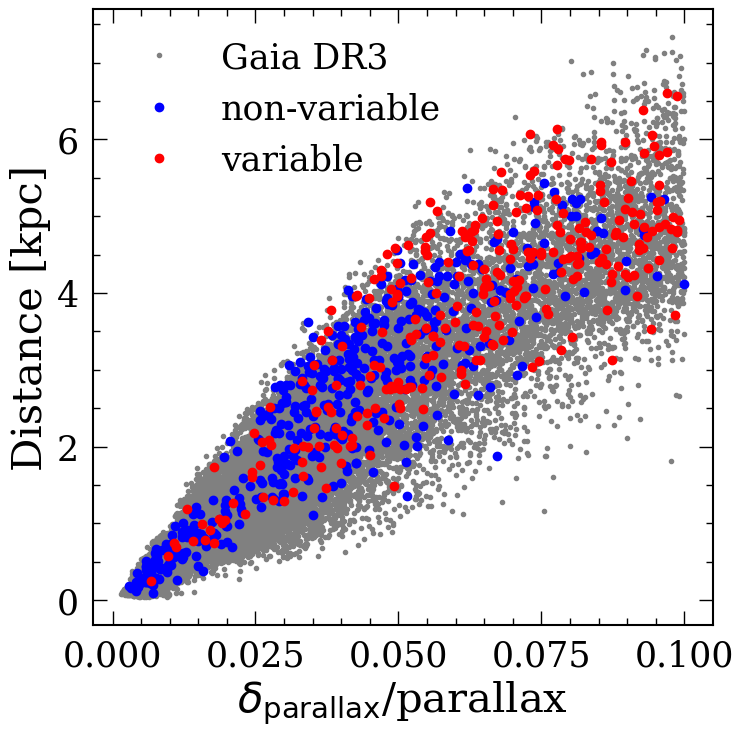}
    \caption{Relative parallax error versus the distance measurements (calculated by 1/parallax) for the full Gaia DR3 sample (grey), the non-variable stars in the RRL instability strip (blue), and the RRLs in the instability strip (red). 
    The non-variable and RRLs exhibit similar uncertainty properties, suggesting incorrect distance measurements (thus incorrect \mg\ measurements) are unlikely.}
    \label{fig:fig3}
\end{figure}

One of the possibilities is some of these non-variable stars happened to land in the instability strip due to incorrect color, magnitude, and distance measurements.
If this is the case, their relative distance uncertainty measurements may exhibit a different characteristic compared to the RRLs that are truly in the instability strip. 
Figure~\ref{fig:fig3} shows the relative parallax error, $\sigma_{\rm parallax}$/parallax, versus distance.
The distance is calculated using 1/parallax. 
It is clear that the 416 non-variable stars (blue) exhibit similar characteristics as the RRLs (red). 
This is not surprising as we only selected stars that have low reddening and excellent parallax measurements.

\subsection{Missing RR Lyrae detection?}\label{subsec:tess}
\begin{figure}
    \centering
    \includegraphics[width=0.95\columnwidth]{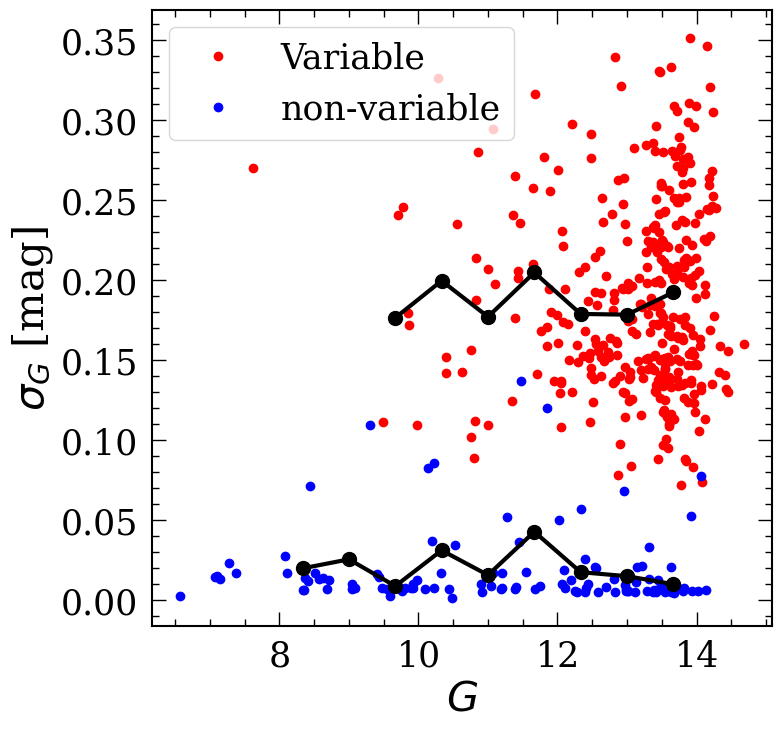}
    \includegraphics[width=0.99\columnwidth]{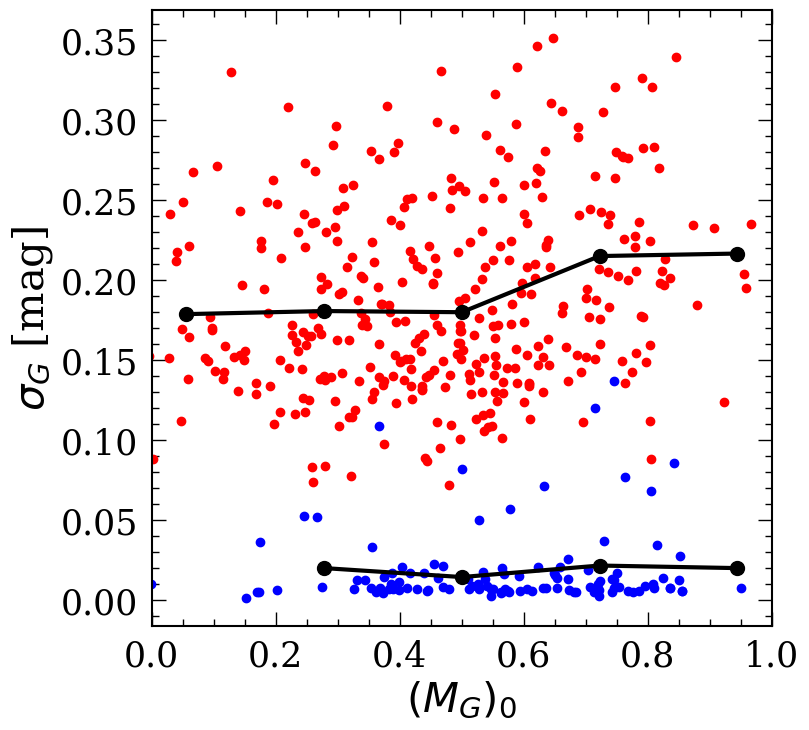}
    \caption{The square root of the unbiased unweighted variance in the $G$--band magnitude for the 112 non-variable stars (blue) and the 360 RRLs (red) as a function of apparently $G$ magnitude (top) and the extinction-corrected absolute $G$ magnitude (bottom).
    These stars occupy the CMD bins where the variability fraction is $>$ 0.2.
    The black lines show the average standard divination for the non-variable stars and the RRLs separately. 
    It is clear that the non-variable stars have significantly lower variation compared to the non-variable stars.
    }
    \label{fig:gaia_var}
\end{figure}

One other possibility is that some RRc or low-amplitude RRLs are not included in the catalogs or misclassification as eclipsing binaries. 
This could happen as it is difficult to vet individual stars by eye for most large RRLs catalogs.
To confirm these stars are truly non-variable, we cross-matched our sample of non-variable and RRLs in \bpgo-\mgo\ with the Gaia DR3 variability catalog (\texttt{gaiadr3.vari\_summary}).
We obtained entries for 112 out of 416 non-variable stars and 360 out of 366 RRLs.
Figure~\ref{fig:gaia_var} shows the square root of the unbiased unweighted variance in the $G$--band magnitude \footnote{column \texttt{std\_dev\_mag\_g\_fov} in the \texttt{gaiadr3.vari\_summary} table}, $\sigma_G$, as a function of apparent $G$ magnitude (top plot) and \mgo\ (bottom plot).
It is obvious that the non-variable stars have significantly lower variance in the $G$--band photometry compared to the identified RRLs, confirming their non-variable nature.
In addition, because we are comparing against the combined Gaia SOS, ASAS-SN, and PS-1 catalog of RRLs, any incompleteness the Gaia SOS catalog may suffer in the identification of RRL as variables is compensated by ASAS+PS1 (at these apparent magnitudes mainly by ASAS), which has many more epochs than Gaia DR3 and, as shown in \citet{Mateu2024}, has better completeness than Gaia SOS in the bright end. 

To further investigate the variability of non-variable stars in the RRL instability strip, we virtually inspected the TESS light curves for the non-variable stars in the CMD bins with variable fraction $>$ 0.5.
We extracted the light curves using the package \texttt{unpopular} \citep{Hattori2022}, an implementation of the Causal Pixel Model de-trending method to obtain TESS Full-Frame Image light curves.
Within the 64 (out of 71) stars with TESS observations, some show extremely low amplitude ($>$ 1\% but mostly $<$ 5\%) variation with periods close to those of RRLs.
For the full set of period-folded light curves, see Figure~\ref{fig:tess}.
However, due to the large pixel size of TESS, it is unclear whether these small amplitude variations we detected are contamination from nearby stars, especially since many of them reside in crowded fields in the Galactic plane. 
Regardless, the light curves of these low-amplitude stars cannot be grouped with the typical RRLs, which should have variations $>$ 0.2 mag.

It is worth pointing out that ultra-low amplitude RRLs have also been reported by \cite{Wallace2019} in the globular cluster M4, varying on the order of 1 millimagnitude.
They suggested these RRLs could be related to the detection of low-amplitude first-overtone modes in other RRLs \citep[e.g.,][]{Gruberbauer2007, Molnar2012, Molnar2022, Netzel2023}, but still lack physical explanation.  
However, these two ultra-amplitude RRLs in M4 are located at the edge of the instability strip, and ours are located at the center of the instability strip.

\subsection{Physical origin?}\label{subsec:phy}

\begin{figure*}
    \centering
    \includegraphics[width=\linewidth]{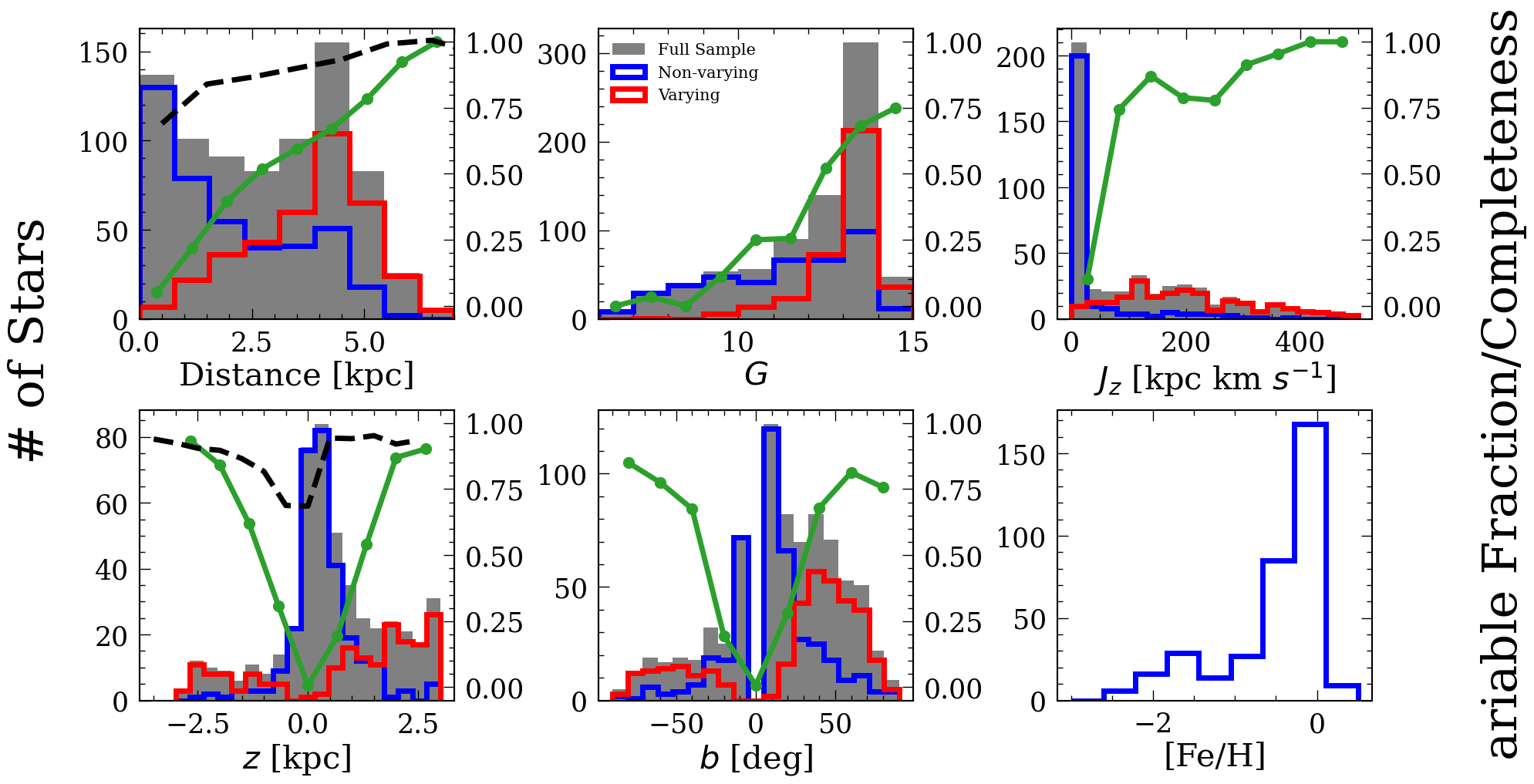}
    \caption{variable fraction (green lines) as a function of distance (top left; calculated using 1/parallax), $G$--band magnitude (top middle), $J_z$ (top right), $z$ (bottom left), and $b$ (bottom middle).
    The bottom right plot shows the metallicity distribution of the non-variable stars with metallicity measurements based on BP/RP spectra \citep{Andrae2023}.
    The black dotted line shows the completeness analysis for RRLs with $|b| > 5$ deg from \cite{Mateu2024}.}
    \label{fig:fig4}
\end{figure*}

Even though it is difficult to fully rule out contamination from other sources, our sample suggests some stars in the RRL instability strip are most likely non-variable.
Typically, RRLs are easy to detect thanks to their large photometric variation amplitude.
However, the TESS light curves suggest some of the closest, brightest stars in the RRL instability strip with low extinction are either not varying or have extremely low amplitude variation. 
In this section, we investigated the variable fraction as a function of distance, $G$--band magnitude, kinematic, and Galactic position. 
To do so, we selected the stars that are in \bpgo--\mgo\ bins with variable fraction $>$ 0.2.
This is the same sample as the blue and red points shown in Figure~\ref{fig:fig3}.
We selected a subgroup of stars that have 6-D kinematic information from Gaia DR3.

The Galactic height, $z$, is calculated from Gaia DR3 \citep{gaiadr3} measurements (RA, Dec., parallax, proper motions, and RV) by transforming from the solar system barycentric ICRS reference frame to the Galactocentric Cartesian and cylindrical coordinates using \texttt{astropy} using updated solar motion parameters from \citet{Hunt:2022}.
The vertical action, $J_z$, is calculated using the \texttt{MilkyWayPotential2022} in \texttt{gala} \citep{gala2017}, with the additional constraint of having a circular velocity at the solar position to be 229 km/s \citep{Eilers2019}. 
We then compute actions using the `St\"{a}ckel Fudge' \citep{Binney2012, Sanders2012} as implemented in \texttt{galpy} \citep{galpy}.
637 stars (321 non-variable and 316 variable) in our sample have full 6-D kinematic information.

Figure~\ref{fig:fig4} shows the variable fraction (green lines) as a function of distance (top left; calculated using 1/parallax), $G$--band magnitude (top middle), $J_z$ (top right), and $z$ (bottom left).
The black dashed lines show the RRLs completeness analysis for stars with $|b| > 5$ deg, using the selection function derived for the Gaia SOS+ASAS+PS1 RRL catalog from \citet{Mateu2024}. 
The selection function was inferred by means of a joint probabilistic analysis of the three surveys, as described in \cite{Mateu2020}. 
This method has the advantage of providing an empirical assessment of the completeness of each of the individual catalogs analyzed, without the assumption of 100\% completeness for any of the catalogs.
It is clear that the non-variable fraction cannot be explained only through the completeness argument.
Moreover, the low completeness towards close-by stars with low Galactic latitude is mostly caused by missing RRc, and we have confirmed using TESS and Gaia that most of the stars in the center of the instability strip are not varying. 
Most non-variable stars in the RRL instability strip are bright ($G$--band magnitude $< \sim$ 12.5 mag), close by ($<$ 5 kpc), and on cold orbits in the Galactic plane ($J_z <$ 50 kpc km s$^{-1}$ and $|z|$ $<$ 1 kpc). 
One may argue what is seen in Figure~\ref{fig:fig4} is contamination from background or foreground stars in the Galactic plane as it is a crowded field. 
However, since we selected stars that are bright and low in extinction, and we are looking for RRL-type variations that should be on the order of several tenths of magnitude, contamination is unlikely.

Since the temperature of the non-variable stars should not fluctuate much, popular spectra fitting or label transfer methods should be able to provide accurate metallicity measurement. 
We cross-matched the sample of non-variable stars with the metallicity catalog from \cite{Andrae2023} and found 354 non-variable stars with metallicity measurements inferred from BP/RP Gaia DR3 spectra.
The metallicity distribution is shown in Figure~\ref{fig:fig4} bottom right plot.
The metallicity of these stars shows a clear peak at [Fe/H] = 0, has a mean metallicity of -0.54 dex with a dispersion of 0.6 dex, agreeing with their kinematic information, and are more metal-rich as a population compared to that of typical RRLs.
This is the opposite of what was predicted in \cite{Cox1973}, as they suggest non-variable stars in the RRL instability strip, if exist, should have lower metallicity and helium content compared to RRLs.
As a result, helium abundance might not be the main cause of non-variable stars in the RRL instability strip. 

The high percentage of non-variable stars in the disk could provide additional information to understand RRLs in the disk, which are metal-rich (although slightly more metal-poor compared to the non-variable stars in our sample) and share the same kinematics characteristic of the intermediate-age population, which is too young to produce RRLs canonically \citep[e.g.,][]{Preston1959, Zinn2020, Prudil2020, Iorio2021}.

\section{Discussion}
There are a few possible explanations for the existence of the non-variable stars in the RRL instability strip.
One possible explanation is these stars are fast rotators. 
Stellar rotation can change the physical structure of stars and break down spherical symmetry \citep[e.g.,][]{Monnier2007}.
This extra dimension can complicate the picture and alter the pulsation frequencies of stars in the instability strip \citep[e.g.,][]{Reese2017, Saio2018}.
In fact, two of the non-variable stars have spectra from the Apache Point Observatory Galactic Evolution Experiment data release 17 \citep[APOGEE DR17;][]{APOGEEDR17}.
$v$sin$i$ measurements suggest these two stars are rotating faster than a typical horizontal branch star.  
However, spectroscopic follow-up is needed to test this idea.

One other possibility is that some non-variable stars are unresolved binaries.
This will alter their measured magnitude as the luminosity of multiple stars are combined into one, making them appear to be in the RRL locus of the CMD. 
Lastly, it is not clear whether a star that is not a horizontal branch star but passes through the RRL instability strip will have the same pulsation properties (or pulsate at all) as RRLs. 
Stars could go through particular evolutionary stages that place them in the same CMD space as RRLs but do not exhibit the same interior structure, and thus, pulsation properties of RRLs. 
Detailed stellar modeling is needed to test this hypothesis.

\section{Conclusion}
We selected a sample of field stars in the RRL instability strip with low extinction and small distance uncertainty using Gaia DR3. 
Using this high-quality sample, we investigated whether non-variable stars exist in the RRL instability strip.
We studied the variable fraction in small bins of \bpg-\mg\ after accounting for extinction with 3D dust maps. 
We chose the bin size so that it is significantly larger than the maximum reddening values inferred for all stars in our sample.
We find that $\sim$ 30\% of stars in the center of the empirical RRL instability strip have significantly smaller photometric variation compared to typical RRLs using Gaia DR3. 
TESS light curves also suggest most of the non-variable stars in bins where the variable fraction is $>$ 0.5 have photometric variation $<$ 5\% based on TESS.
These non-variable stars are mostly bright and close by, on cold orbits in the Galactic plane.
Metallicity from Gaia BP/RP spectra suggests their average metallicity is $\sim$ -1 dex, with a peak at 0. 
The discovery of these non-variable stars in the RRL instability strip challenges our understanding of stellar physics and further investigation such as spectroscopy or asteroseismology follow-up is needed to understand the origin of these stars. 

\begin{acknowledgments}
\section{Acknowledgements}
We want to thank the participants of the morning ``Astronomy Coffee'' at the Department of Astronomy, The Ohio State University, for the daily and lively astro-ph discussion, one of which prompted us to investigate the problem described in this paper. 
The authors thank the helpful discussion with Marc H. Pinsonneault and Jennifer A. Johnson. 
The authors thank Adrian Price-Whelan for producing the kinematic data using Gaia EDR3 measurements. 
This work has made use of data from the European Space Agency (ESA) mission Gaia,\footnote{\url{https://www.cosmos.esa.int/gaia}} processed by the Gaia Data Processing and Analysis Consortium (DPAC).\footnote{\url{https://www.cosmos.esa.int/web/gaia/dpac/consortium}} Funding for the DPAC has been provided by national institutions, in particular the institutions participating in the Gaia Multilateral Agreement.
This research also made use of public auxiliary data provided by ESA/Gaia/DPAC/CU5 and prepared by Carine Babusiaux. 
This research has also made use of NASA's Astrophysics Data System, 
and the VizieR \citep{vizier} and SIMBAD \citep{simbad} databases, 
operated at CDS, Strasbourg, France.
This research or product makes use of public auxiliary data provided by ESA/Gaia/DPAC/CU5 and prepared by Carine Babusiaux
This research has been supported by funding from the project FCE\_1\_2021\_1\_167524, Fondo Clemente Estable, Agencia Nacional de Innovaci\'on e Investigaci\'on (ANII). 
KZS is supported by NSF grants AST-2307385 and 2407206.
\end{acknowledgments}

%

\vspace{5mm}
\facilities{Gaia, TESS, APOGEE}


\software{\texttt{Astropy} \citep{astropy:2013, astropy:2018, astropy2022},
\texttt{lightkurve} \citep{lightkurve2018},
            \texttt{Matplotlib} \citep{matplotlib}, 
            \texttt{NumPy} \citep{Numpy}, 
            \texttt{Pandas} \citep{pandas},
            \texttt{dustmaps} \citep{Green2018},
            \texttt{unpopular} \citep{Hattori2022}}
\\
\\
\\
\appendix
\renewcommand\thefigure{\thesection.\arabic{figure}}    
\section{Appendix}
\setcounter{figure}{0}    
TESS light curves extracted using the package \texttt{unpopular} \citep{Hattori2022}, an implementation of the Causal Pixel Model de-trending method to obtain TESS Full-Frame Image light curves.
The periods were then detected using the \texttt{astropy} \citep{astropy:2013, astropy:2018, astropy2022} implementation of the Lomb--Scargle periodogram. 
Figure~\ref{fig:tess} shows the period--folded light curves for 64 out of 71 stars, and the Gaia DR3 ID for the 4 stars without TESS light curves are 4097551529423348864, 4094755436991875072, 2941935847667837696, and 4069920497296406272. 
\begin{figure*}
    \centering
    \includegraphics[width=\linewidth]{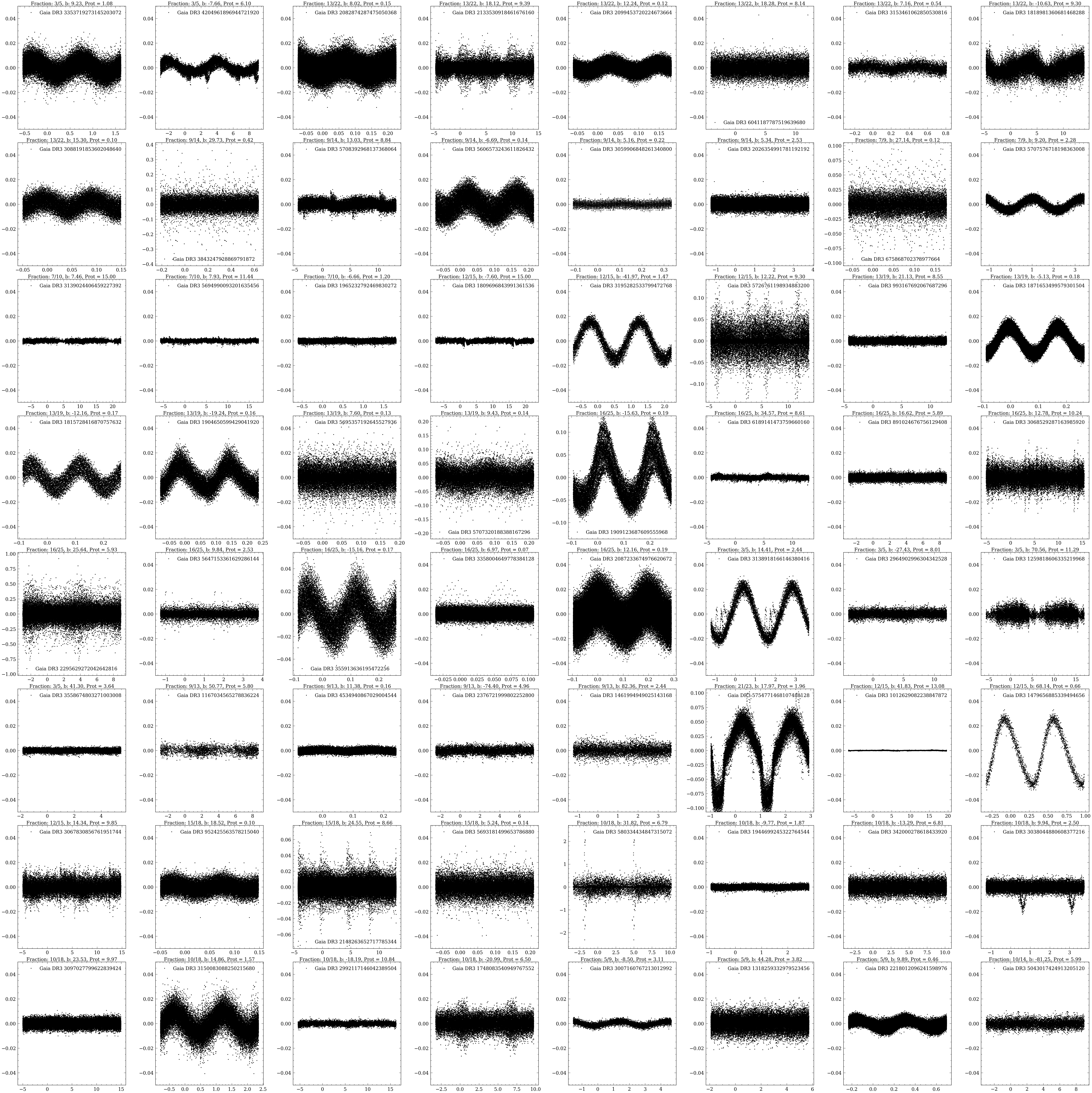}
    \caption{Period folded TESS light curves for 64 out of 71 non-variable stars in the \bpgo--\mgo\ bins where the variable fraction is $>$ 0.5, generated using the \texttt{unpopular} package \citep{Hattori2022}.
    For each star, the folded light curves are shown twice for better visualization. 
    The title shows the number of variable stars/the total number of stars in the \bpgo--\mgo\ bin where that star belongs as well as the detected period.
    The $y$-axis of each subplot shows the normalized flux.
    For most stars, the light curve variability is significantly smaller than what is expected for RRLs, suggesting they are likely not RRLs.}
    \label{fig:tess}
\end{figure*}

\bibliography{sample631}{}

\begin{thebibliography}{}
\expandafter\ifx\csname natexlab\endcsname\relax\def\natexlab#1{#1}\fi
\providecommand{\url}[1]{\href{#1}{#1}}
\providecommand{\dodoi}[1]{doi:~\href{http://doi.org/#1}{\nolinkurl{#1}}}
\providecommand{\doeprint}[1]{\href{http://ascl.net/#1}{\nolinkurl{http://ascl.net/#1}}}
\providecommand{\doarXiv}[1]{\href{https://arxiv.org/abs/#1}{\nolinkurl{https://arxiv.org/abs/#1}}}

\bibitem[{{Abdurro'uf} {et~al.}(2022){Abdurro'uf}, {Accetta}, {Aerts}, {Silva Aguirre}, {Ahumada}, {Ajgaonkar}, {Filiz Ak}, {Alam}, {Allende Prieto}, {Almeida}, {Anders}, {Anderson}, {Andrews}, {Anguiano}, {Aquino-Ort{\'\i}z}, {Arag{\'o}n-Salamanca}, {Argudo-Fern{\'a}ndez}, {Ata}, {Aubert}, {Avila-Reese}, {Badenes}, {Barb{\'a}}, {Barger}, {Barrera-Ballesteros}, {Beaton}, {Beers}, {Belfiore}, {Bender}, {Bernardi}, {Bershady}, {Beutler}, {Bidin}, {Bird}, {Bizyaev}, {Blanc}, {Blanton}, {Boardman}, {Bolton}, {Boquien}, {Borissova}, {Bovy}, {Brandt}, {Brown}, {Brownstein}, {Brusa}, {Buchner}, {Bundy}, {Burchett}, {Bureau}, {Burgasser}, {Cabang}, {Campbell}, {Cappellari}, {Carlberg}, {Wanderley}, {Carrera}, {Cash}, {Chen}, {Chen}, {Cherinka}, {Chiappini}, {Choi}, {Chojnowski}, {Chung}, {Clerc}, {Cohen}, {Comerford}, {Comparat}, {da Costa}, {Covey}, {Crane}, {Cruz-Gonzalez}, {Culhane}, {Cunha}, {Dai}, {Damke}, {Darling}, {Davidson}, {Davies}, {Dawson}, {De Lee}, {Diamond-Stanic}, {Cano-D{\'\i}az}, {S{\'a}nchez},
  {Donor}, {Duckworth}, {Dwelly}, {Eisenstein}, {Elsworth}, {Emsellem}, {Eracleous}, {Escoffier}, {Fan}, {Farr}, {Feng}, {Fern{\'a}ndez-Trincado}, {Feuillet}, {Filipp}, {Fillingham}, {Frinchaboy}, {Fromenteau}, {Galbany}, {Garc{\'\i}a}, {Garc{\'\i}a-Hern{\'a}ndez}, {Ge}, {Geisler}, {Gelfand}, {G{\'e}ron}, {Gibson}, {Goddy}, {Godoy-Rivera}, {Grabowski}, {Green}, {Greener}, {Grier}, {Griffith}, {Guo}, {Guy}, {Hadjara}, {Harding}, {Hasselquist}, {Hayes}, {Hearty}, {Hern{\'a}ndez}, {Hill}, {Hogg}, {Holtzman}, {Horta}, {Hsieh}, {Hsu}, {Hsu}, {Huber}, {Huertas-Company}, {Hutchinson}, {Hwang}, {Ibarra-Medel}, {Chitham}, {Ilha}, {Imig}, {Jaekle}, {Jayasinghe}, {Ji}, {Johnson}, {Jones}, {J{\"o}nsson}, {Katkov}, {Khalatyan}, {Kinemuchi}, {Kisku}, {Knapen}, {Kneib}, {Kollmeier}, {Kong}, {Kounkel}, {Kreckel}, {Krishnarao}, {Lacerna}, {Lane}, {Langgin}, {Lavender}, {Law}, {Lazarz}, {Leung}, {Leung}, {Lewis}, {Li}, {Li}, {Lian}, {Liang}, {Lin}, {Lin}, {Lin}, {Lintott}, {Long}, {Longa-Pe{\~n}a}, {L{\'o}pez-Cob{\'a}}, {Lu},
  {Lundgren}, {Luo}, {Mackereth}, {de la Macorra}, {Mahadevan}, {Majewski}, {Manchado}, {Mandeville}, {Maraston}, {Margalef-Bentabol}, {Masseron}, {Masters}, {Mathur}, {McDermid}, {Mckay}, {Merloni}, {Merrifield}, {Meszaros}, {Miglio}, {Di Mille}, {Minniti}, {Minsley}, {Monachesi}, {Moon}, {Mosser}, {Mulchaey}, {Muna}, {Mu{\~n}oz}, {Myers}, {Myers}, {Nadathur}, {Nair}, {Nandra}, {Neumann}, {Newman}, {Nidever}, {Nikakhtar}, {Nitschelm}, {O'Connell}, {Garma-Oehmichen}, {Luan Souza de Oliveira}, {Olney}, {Oravetz}, {Ortigoza-Urdaneta}, {Osorio}, {Otter}, {Pace}, {Padilla}, {Pan}, {Pan}, {Parikh}, {Parker}, {Peirani}, {Pe{\~n}a Ram{\'\i}rez}, {Penny}, {Percival}, {Perez-Fournon}, {Pinsonneault}, {Poidevin}, {Poovelil}, {Price-Whelan}, {B{\'a}rbara de Andrade Queiroz}, {Raddick}, {Ray}, {Rembold}, {Riddle}, {Riffel}, {Riffel}, {Rix}, {Robin}, {Rodr{\'\i}guez-Puebla}, {Roman-Lopes}, {Rom{\'a}n-Z{\'u}{\~n}iga}, {Rose}, {Ross}, {Rossi}, {Rubin}, {Salvato}, {S{\'a}nchez}, {S{\'a}nchez-Gallego}, {Sanderson}, {Santana
  Rojas}, {Sarceno}, {Sarmiento}, {Sayres}, {Sazonova}, {Schaefer}, {Schiavon}, {Schlegel}, {Schneider}, {Schultheis}, {Schwope}, {Serenelli}, {Serna}, {Shao}, {Shapiro}, {Sharma}, {Shen}, {Shetrone}, {Shu}, {Simon}, {Skrutskie}, {Smethurst}, {Smith}, {Sobeck}, {Spoo}, {Sprague}, {Stark}, {Stassun}, {Steinmetz}, {Stello}, {Stone-Martinez}, {Storchi-Bergmann}, {Stringfellow}, {Stutz}, {Su}, {Taghizadeh-Popp}, {Talbot}, {Tayar}, {Telles}, {Teske}, {Thakar}, {Theissen}, {Tkachenko}, {Thomas}, {Tojeiro}, {Hernandez Toledo}, {Troup}, {Trump}, {Trussler}, {Turner}, {Tuttle}, {Unda-Sanzana}, {V{\'a}zquez-Mata}, {Valentini}, {Valenzuela}, {Vargas-Gonz{\'a}lez}, {Vargas-Maga{\~n}a}, {Alfaro}, {Villanova}, {Vincenzo}, {Wake}, {Warfield}, {Washington}, {Weaver}, {Weijmans}, {Weinberg}, {Weiss}, {Westfall}, {Wild}, {Wilde}, {Wilson}, {Wilson}, {Wilson}, {Wolf}, {Wood-Vasey}, {Yan}, {Zamora}, {Zasowski}, {Zhang}, {Zhao}, {Zheng}, {Zheng}, \& {Zhu}}]{APOGEEDR17}
{Abdurro'uf}, {Accetta}, K., {Aerts}, C., {et~al.} 2022, \apjs, 259, 35, \dodoi{10.3847/1538-4365/ac4414}

\bibitem[{{Andrae} {et~al.}(2023){Andrae}, {Rix}, \& {Chandra}}]{Andrae2023}
{Andrae}, R., {Rix}, H.-W., \& {Chandra}, V. 2023, \apjs, 267, 8, \dodoi{10.3847/1538-4365/acd53e}

\bibitem[{{Andrievsky} \& {Kovtyukh}(1996)}]{Andrievsky1996}
{Andrievsky}, S.~M., \& {Kovtyukh}, V.~V. 1996, \apss, 245, 61, \dodoi{10.1007/BF00637803}

\bibitem[{{Astropy Collaboration} {et~al.}(2013){Astropy Collaboration}, {Robitaille}, {Tollerud}, {Greenfield}, {Droettboom}, {Bray}, {Aldcroft}, {Davis}, {Ginsburg}, {Price-Whelan}, {Kerzendorf}, {Conley}, {Crighton}, {Barbary}, {Muna}, {Ferguson}, {Grollier}, {Parikh}, {Nair}, {Unther}, {Deil}, {Woillez}, {Conseil}, {Kramer}, {Turner}, {Singer}, {Fox}, {Weaver}, {Zabalza}, {Edwards}, {Azalee Bostroem}, {Burke}, {Casey}, {Crawford}, {Dencheva}, {Ely}, {Jenness}, {Labrie}, {Lim}, {Pierfederici}, {Pontzen}, {Ptak}, {Refsdal}, {Servillat}, \& {Streicher}}]{astropy:2013}
{Astropy Collaboration}, {Robitaille}, T.~P., {Tollerud}, E.~J., {et~al.} 2013, \aap, 558, A33, \dodoi{10.1051/0004-6361/201322068}

\bibitem[{{Astropy Collaboration} {et~al.}(2022){Astropy Collaboration}, {Price-Whelan}, {Lim}, {Earl}, {Starkman}, {Bradley}, {Shupe}, {Patil}, {Corrales}, {Brasseur}, {N{\"o}the}, {Donath}, {Tollerud}, {Morris}, {Ginsburg}, {Vaher}, {Weaver}, {Tocknell}, {Jamieson}, {van Kerkwijk}, {Robitaille}, {Merry}, {Bachetti}, {G{\"u}nther}, {Aldcroft}, {Alvarado-Montes}, {Archibald}, {B{\'o}di}, {Bapat}, {Barentsen}, {Baz{\'a}n}, {Biswas}, {Boquien}, {Burke}, {Cara}, {Cara}, {Conroy}, {Conseil}, {Craig}, {Cross}, {Cruz}, {D'Eugenio}, {Dencheva}, {Devillepoix}, {Dietrich}, {Eigenbrot}, {Erben}, {Ferreira}, {Foreman-Mackey}, {Fox}, {Freij}, {Garg}, {Geda}, {Glattly}, {Gondhalekar}, {Gordon}, {Grant}, {Greenfield}, {Groener}, {Guest}, {Gurovich}, {Handberg}, {Hart}, {Hatfield-Dodds}, {Homeier}, {Hosseinzadeh}, {Jenness}, {Jones}, {Joseph}, {Kalmbach}, {Karamehmetoglu}, {Ka{\l}uszy{\'n}ski}, {Kelley}, {Kern}, {Kerzendorf}, {Koch}, {Kulumani}, {Lee}, {Ly}, {Ma}, {MacBride}, {Maljaars}, {Muna}, {Murphy}, {Norman},
  {O'Steen}, {Oman}, {Pacifici}, {Pascual}, {Pascual-Granado}, {Patil}, {Perren}, {Pickering}, {Rastogi}, {Roulston}, {Ryan}, {Rykoff}, {Sabater}, {Sakurikar}, {Salgado}, {Sanghi}, {Saunders}, {Savchenko}, {Schwardt}, {Seifert-Eckert}, {Shih}, {Jain}, {Shukla}, {Sick}, {Simpson}, {Singanamalla}, {Singer}, {Singhal}, {Sinha}, {Sip{\H{o}}cz}, {Spitler}, {Stansby}, {Streicher}, {{\v{S}}umak}, {Swinbank}, {Taranu}, {Tewary}, {Tremblay}, {de Val-Borro}, {Van Kooten}, {Vasovi{\'c}}, {Verma}, {de Miranda Cardoso}, {Williams}, {Wilson}, {Winkel}, {Wood-Vasey}, {Xue}, {Yoachim}, {Zhang}, {Zonca}, \& {Astropy Project Contributors}}]{astropy2022}
{Astropy Collaboration}, {Price-Whelan}, A.~M., {Lim}, P.~L., {et~al.} 2022, \apj, 935, 167, \dodoi{10.3847/1538-4357/ac7c74}

\bibitem[{{Bellm} {et~al.}(2019){Bellm}, {Kulkarni}, {Graham}, {Dekany}, {Smith}, {Riddle}, {Masci}, {Helou}, {Prince}, {Adams}, {Barbarino}, {Barlow}, {Bauer}, {Beck}, {Belicki}, {Biswas}, {Blagorodnova}, {Bodewits}, {Bolin}, {Brinnel}, {Brooke}, {Bue}, {Bulla}, {Burruss}, {Cenko}, {Chang}, {Connolly}, {Coughlin}, {Cromer}, {Cunningham}, {De}, {Delacroix}, {Desai}, {Duev}, {Eadie}, {Farnham}, {Feeney}, {Feindt}, {Flynn}, {Franckowiak}, {Frederick}, {Fremling}, {Gal-Yam}, {Gezari}, {Giomi}, {Goldstein}, {Golkhou}, {Goobar}, {Groom}, {Hacopians}, {Hale}, {Henning}, {Ho}, {Hover}, {Howell}, {Hung}, {Huppenkothen}, {Imel}, {Ip}, {Ivezi{\'c}}, {Jackson}, {Jones}, {Juric}, {Kasliwal}, {Kaspi}, {Kaye}, {Kelley}, {Kowalski}, {Kramer}, {Kupfer}, {Landry}, {Laher}, {Lee}, {Lin}, {Lin}, {Lunnan}, {Giomi}, {Mahabal}, {Mao}, {Miller}, {Monkewitz}, {Murphy}, {Ngeow}, {Nordin}, {Nugent}, {Ofek}, {Patterson}, {Penprase}, {Porter}, {Rauch}, {Rebbapragada}, {Reiley}, {Rigault}, {Rodriguez}, {van Roestel}, {Rusholme}, {van
  Santen}, {Schulze}, {Shupe}, {Singer}, {Soumagnac}, {Stein}, {Surace}, {Sollerman}, {Szkody}, {Taddia}, {Terek}, {Van Sistine}, {van Velzen}, {Vestrand}, {Walters}, {Ward}, {Ye}, {Yu}, {Yan}, \& {Zolkower}}]{ZTF}
{Bellm}, E.~C., {Kulkarni}, S.~R., {Graham}, M.~J., {et~al.} 2019, \pasp, 131, 018002, \dodoi{10.1088/1538-3873/aaecbe}

\bibitem[{{Binney}(2012)}]{Binney2012}
{Binney}, J. 2012, \mnras, 426, 1324, \dodoi{10.1111/j.1365-2966.2012.21757.x}

\bibitem[{{Bono}(2003)}]{Bono2003}
{Bono}, G. 2003, in Stellar Candles for the Extragalactic Distance Scale, ed. D.~{Alloin} \& W.~{Gieren}, Vol. 635, 85--104, \dodoi{10.1007/978-3-540-39882-0_5}

\bibitem[{{Borucki} {et~al.}(2010){Borucki}, {Koch}, {Basri}, {Batalha}, {Brown}, {Caldwell}, {Caldwell}, {Christensen-Dalsgaard}, {Cochran}, {DeVore}, {Dunham}, {Dupree}, {Gautier}, {Geary}, {Gilliland}, {Gould}, {Howell}, {Jenkins}, {Kondo}, {Latham}, {Marcy}, {Meibom}, {Kjeldsen}, {Lissauer}, {Monet}, {Morrison}, {Sasselov}, {Tarter}, {Boss}, {Brownlee}, {Owen}, {Buzasi}, {Charbonneau}, {Doyle}, {Fortney}, {Ford}, {Holman}, {Seager}, {Steffen}, {Welsh}, {Rowe}, {Anderson}, {Buchhave}, {Ciardi}, {Walkowicz}, {Sherry}, {Horch}, {Isaacson}, {Everett}, {Fischer}, {Torres}, {Johnson}, {Endl}, {MacQueen}, {Bryson}, {Dotson}, {Haas}, {Kolodziejczak}, {Van Cleve}, {Chandrasekaran}, {Twicken}, {Quintana}, {Clarke}, {Allen}, {Li}, {Wu}, {Tenenbaum}, {Verner}, {Bruhweiler}, {Barnes}, \& {Prsa}}]{kepler}
{Borucki}, W.~J., {Koch}, D., {Basri}, G., {et~al.} 2010, Science, 327, 977, \dodoi{10.1126/science.1185402}

\bibitem[{{Bovy}(2015)}]{galpy}
{Bovy}, J. 2015, \apjs, 216, 29, \dodoi{10.1088/0067-0049/216/2/29}

\bibitem[{{Cabrera-Gadea} {et~al.}(2024){Cabrera-Gadea}, {Mateu}, \& {Ramos}}]{CabreraGadea2024b}
{Cabrera-Gadea}, M., {Mateu}, C., \& {Ramos}, P. 2024, arXiv e-prints, arXiv:2410.22427.
\newblock \doarXiv{2410.22427}

\bibitem[{{Caputo} {et~al.}(2000){Caputo}, {Castellani}, {Marconi}, \& {Ripepi}}]{Caputo2000}
{Caputo}, F., {Castellani}, V., {Marconi}, M., \& {Ripepi}, V. 2000, \mnras, 316, 819, \dodoi{10.1046/j.1365-8711.2000.03591.x}

\bibitem[{{Clementini} {et~al.}(2023){Clementini}, {Ripepi}, {Garofalo}, {Molinaro}, {Muraveva}, {Leccia}, {Rimoldini}, {Holl}, {Jevardat de Fombelle}, {Sartoretti}, {Marchal}, {Audard}, {Nienartowicz}, {Andrae}, {Marconi}, {Szabados}, {Evans}, {Lecoeur-Taibi}, {Mowlavi}, {Musella}, \& {Eyer}}]{Clementini2023}
{Clementini}, G., {Ripepi}, V., {Garofalo}, A., {et~al.} 2023, \aap, 674, A18, \dodoi{10.1051/0004-6361/202243964}

\bibitem[{{Cox} {et~al.}(1973){Cox}, {King}, \& {Tabor}}]{Cox1973}
{Cox}, A.~N., {King}, D.~S., \& {Tabor}, J.~E. 1973, \apj, 184, 201, \dodoi{10.1086/152319}

\bibitem[{{Cox}(1963)}]{Cox1963}
{Cox}, J.~P. 1963, \apj, 138, 487, \dodoi{10.1086/147661}

\bibitem[{{Cruz Reyes} {et~al.}(2024){Cruz Reyes}, {Anderson}, {Johansson}, {Netzel}, \& {Medaric}}]{CruzReyes2024}
{Cruz Reyes}, M., {Anderson}, R.~I., {Johansson}, L., {Netzel}, H., \& {Medaric}, Z. 2024, \aap, 684, A173, \dodoi{10.1051/0004-6361/202348961}

\bibitem[{{Danielski} {et~al.}(2018){Danielski}, {Babusiaux}, {Ruiz-Dern}, {Sartoretti}, \& {Arenou}}]{Danielski2018}
{Danielski}, C., {Babusiaux}, C., {Ruiz-Dern}, L., {Sartoretti}, P., \& {Arenou}, F. 2018, \aap, 614, A19, \dodoi{10.1051/0004-6361/201732327}

\bibitem[{{Eilers} {et~al.}(2019){Eilers}, {Hogg}, {Rix}, \& {Ness}}]{Eilers2019}
{Eilers}, A.-C., {Hogg}, D.~W., {Rix}, H.-W., \& {Ness}, M.~K. 2019, \apj, 871, 120, \dodoi{10.3847/1538-4357/aaf648}

\bibitem[{{Gaia Collaboration} {et~al.}(2016){Gaia Collaboration}, {Prusti}, {de Bruijne}, {Brown}, {Vallenari}, {Babusiaux}, {Bailer-Jones}, {Bastian}, {Biermann}, {Evans}, {Eyer}, {Jansen}, {Jordi}, {Klioner}, {Lammers}, {Lindegren}, {Luri}, {Mignard}, {Milligan}, {Panem}, {Poinsignon}, {Pourbaix}, {Randich}, {Sarri}, {Sartoretti}, {Siddiqui}, {Soubiran}, {Valette}, {van Leeuwen}, {Walton}, {Aerts}, {Arenou}, {Cropper}, {Drimmel}, {H{\o}g}, {Katz}, {Lattanzi}, {O'Mullane}, {Grebel}, {Holland}, {Huc}, {Passot}, {Bramante}, {Cacciari}, {Casta{\~n}eda}, {Chaoul}, {Cheek}, {De Angeli}, {Fabricius}, {Guerra}, {Hern{\'a}ndez}, {Jean-Antoine-Piccolo}, {Masana}, {Messineo}, {Mowlavi}, {Nienartowicz}, {Ord{\'o}{\~n}ez-Blanco}, {Panuzzo}, {Portell}, {Richards}, {Riello}, {Seabroke}, {Tanga}, {Th{\'e}venin}, {Torra}, {Els}, {Gracia-Abril}, {Comoretto}, {Garcia-Reinaldos}, {Lock}, {Mercier}, {Altmann}, {Andrae}, {Astraatmadja}, {Bellas-Velidis}, {Benson}, {Berthier}, {Blomme}, {Busso}, {Carry}, {Cellino}, {Clementini},
  {Cowell}, {Creevey}, {Cuypers}, {Davidson}, {De Ridder}, {de Torres}, {Delchambre}, {Dell'Oro}, {Ducourant}, {Fr{\'e}mat}, {Garc{\'\i}a-Torres}, {Gosset}, {Halbwachs}, {Hambly}, {Harrison}, {Hauser}, {Hestroffer}, {Hodgkin}, {Huckle}, {Hutton}, {Jasniewicz}, {Jordan}, {Kontizas}, {Korn}, {Lanzafame}, {Manteiga}, {Moitinho}, {Muinonen}, {Osinde}, {Pancino}, {Pauwels}, {Petit}, {Recio-Blanco}, {Robin}, {Sarro}, {Siopis}, {Smith}, {Smith}, {Sozzetti}, {Thuillot}, {van Reeven}, {Viala}, {Abbas}, {Abreu Aramburu}, {Accart}, {Aguado}, {Allan}, {Allasia}, {Altavilla}, {{\'A}lvarez}, {Alves}, {Anderson}, {Andrei}, {Anglada Varela}, {Antiche}, {Antoja}, {Ant{\'o}n}, {Arcay}, {Atzei}, {Ayache}, {Bach}, {Baker}, {Balaguer-N{\'u}{\~n}ez}, {Barache}, {Barata}, {Barbier}, {Barblan}, {Baroni}, {Barrado y Navascu{\'e}s}, {Barros}, {Barstow}, {Becciani}, {Bellazzini}, {Bellei}, {Bello Garc{\'\i}a}, {Belokurov}, {Bendjoya}, {Berihuete}, {Bianchi}, {Bienaym{\'e}}, {Billebaud}, {Blagorodnova}, {Blanco-Cuaresma}, {Boch},
  {Bombrun}, {Borrachero}, {Bouquillon}, {Bourda}, {Bouy}, {Bragaglia}, {Breddels}, {Brouillet}, {Br{\"u}semeister}, {Bucciarelli}, {Budnik}, {Burgess}, {Burgon}, {Burlacu}, {Busonero}, {Buzzi}, {Caffau}, {Cambras}, {Campbell}, {Cancelliere}, {Cantat-Gaudin}, {Carlucci}, {Carrasco}, {Castellani}, {Charlot}, {Charnas}, {Charvet}, {Chassat}, {Chiavassa}, {Clotet}, {Cocozza}, {Collins}, {Collins}, {Costigan}, {Crifo}, {Cross}, {Crosta}, {Crowley}, {Dafonte}, {Damerdji}, {Dapergolas}, {David}, {David}, {De Cat}, {de Felice}, {de Laverny}, {De Luise}, {De March}, {de Martino}, {de Souza}, {Debosscher}, {del Pozo}, {Delbo}, {Delgado}, {Delgado}, {di Marco}, {Di Matteo}, {Diakite}, {Distefano}, {Dolding}, {Dos Anjos}, {Drazinos}, {Dur{\'a}n}, {Dzigan}, {Ecale}, {Edvardsson}, {Enke}, {Erdmann}, {Escolar}, {Espina}, {Evans}, {Eynard Bontemps}, {Fabre}, {Fabrizio}, {Faigler}, {Falc{\~a}o}, {Farr{\`a}s Casas}, {Faye}, {Federici}, {Fedorets}, {Fern{\'a}ndez-Hern{\'a}ndez}, {Fernique}, {Fienga}, {Figueras}, {Filippi},
  {Findeisen}, {Fonti}, {Fouesneau}, {Fraile}, {Fraser}, {Fuchs}, {Furnell}, {Gai}, {Galleti}, {Galluccio}, {Garabato}, {Garc{\'\i}a-Sedano}, {Gar{\'e}}, {Garofalo}, {Garralda}, {Gavras}, {Gerssen}, {Geyer}, {Gilmore}, {Girona}, {Giuffrida}, {Gomes}, {Gonz{\'a}lez-Marcos}, {Gonz{\'a}lez-N{\'u}{\~n}ez}, {Gonz{\'a}lez-Vidal}, {Granvik}, {Guerrier}, {Guillout}, {Guiraud}, {G{\'u}rpide}, {Guti{\'e}rrez-S{\'a}nchez}, {Guy}, {Haigron}, {Hatzidimitriou}, {Haywood}, {Heiter}, {Helmi}, {Hobbs}, {Hofmann}, {Holl}, {Holland}, {Hunt}, {Hypki}, {Icardi}, {Irwin}, {Jevardat de Fombelle}, {Jofr{\'e}}, {Jonker}, {Jorissen}, {Julbe}, {Karampelas}, {Kochoska}, {Kohley}, {Kolenberg}, {Kontizas}, {Koposov}, {Kordopatis}, {Koubsky}, {Kowalczyk}, {Krone-Martins}, {Kudryashova}, {Kull}, {Bachchan}, {Lacoste-Seris}, {Lanza}, {Lavigne}, {Le Poncin-Lafitte}, {Lebreton}, {Lebzelter}, {Leccia}, {Leclerc}, {Lecoeur-Taibi}, {Lemaitre}, {Lenhardt}, {Leroux}, {Liao}, {Licata}, {Lindstr{\o}m}, {Lister}, {Livanou}, {Lobel}, {L{\"o}ffler},
  {L{\'o}pez}, {Lopez-Lozano}, {Lorenz}, {Loureiro}, {MacDonald}, {Magalh{\~a}es Fernandes}, {Managau}, {Mann}, {Mantelet}, {Marchal}, {Marchant}, {Marconi}, {Marie}, {Marinoni}, {Marrese}, {Marschalk{\'o}}, {Marshall}, {Mart{\'\i}n-Fleitas}, {Martino}, {Mary}, {Matijevi{\v{c}}}, {Mazeh}, {McMillan}, {Messina}, {Mestre}, {Michalik}, {Millar}, {Miranda}, {Molina}, {Molinaro}, {Molinaro}, {Moln{\'a}r}, {Moniez}, {Montegriffo}, {Monteiro}, {Mor}, {Mora}, {Morbidelli}, {Morel}, {Morgenthaler}, {Morley}, {Morris}, {Mulone}, {Muraveva}, {Musella}, {Narbonne}, {Nelemans}, {Nicastro}, {Noval}, {Ord{\'e}novic}, {Ordieres-Mer{\'e}}, {Osborne}, {Pagani}, {Pagano}, {Pailler}, {Palacin}, {Palaversa}, {Parsons}, {Paulsen}, {Pecoraro}, {Pedrosa}, {Pentik{\"a}inen}, {Pereira}, {Pichon}, {Piersimoni}, {Pineau}, {Plachy}, {Plum}, {Poujoulet}, {Pr{\v{s}}a}, {Pulone}, {Ragaini}, {Rago}, {Rambaux}, {Ramos-Lerate}, {Ranalli}, {Rauw}, {Read}, {Regibo}, {Renk}, {Reyl{\'e}}, {Ribeiro}, {Rimoldini}, {Ripepi}, {Riva}, {Rixon},
  {Roelens}, {Romero-G{\'o}mez}, {Rowell}, {Royer}, {Rudolph}, {Ruiz-Dern}, {Sadowski}, {Sagrist{\`a} Sell{\'e}s}, {Sahlmann}, {Salgado}, {Salguero}, {Sarasso}, {Savietto}, {Schnorhk}, {Schultheis}, {Sciacca}, {Segol}, {Segovia}, {Segransan}, {Serpell}, {Shih}, {Smareglia}, {Smart}, {Smith}, {Solano}, {Solitro}, {Sordo}, {Soria Nieto}, {Souchay}, {Spagna}, {Spoto}, {Stampa}, {Steele}, {Steidelm{\"u}ller}, {Stephenson}, {Stoev}, {Suess}, {S{\"u}veges}, {Surdej}, {Szabados}, {Szegedi-Elek}, {Tapiador}, {Taris}, {Tauran}, {Taylor}, {Teixeira}, {Terrett}, {Tingley}, {Trager}, {Turon}, {Ulla}, {Utrilla}, {Valentini}, {van Elteren}, {Van Hemelryck}, {van Leeuwen}, {Varadi}, {Vecchiato}, {Veljanoski}, {Via}, {Vicente}, {Vogt}, {Voss}, {Votruba}, {Voutsinas}, {Walmsley}, {Weiler}, {Weingrill}, {Werner}, {Wevers}, {Whitehead}, {Wyrzykowski}, {Yoldas}, {{\v{Z}}erjal}, {Zucker}, {Zurbach}, {Zwitter}, {Alecu}, {Allen}, {Allende Prieto}, {Amorim}, {Anglada-Escud{\'e}}, {Arsenijevic}, {Azaz}, {Balm}, {Beck}, {Bernstein},
  {Bigot}, {Bijaoui}, {Blasco}, {Bonfigli}, {Bono}, {Boudreault}, {Bressan}, {Brown}, {Brunet}, {Bunclark}, {Buonanno}, {Butkevich}, {Carret}, {Carrion}, {Chemin}, {Ch{\'e}reau}, {Corcione}, {Darmigny}, {de Boer}, {de Teodoro}, {de Zeeuw}, {Delle Luche}, {Domingues}, {Dubath}, {Fodor}, {Fr{\'e}zouls}, {Fries}, {Fustes}, {Fyfe}, {Gallardo}, {Gallegos}, {Gardiol}, {Gebran}, {Gomboc}, {G{\'o}mez}, {Grux}, {Gueguen}, {Heyrovsky}, {Hoar}, {Iannicola}, {Isasi Parache}, {Janotto}, {Joliet}, {Jonckheere}, {Keil}, {Kim}, {Klagyivik}, {Klar}, {Knude}, {Kochukhov}, {Kolka}, {Kos}, {Kutka}, {Lainey}, {LeBouquin}, {Liu}, {Loreggia}, {Makarov}, {Marseille}, {Martayan}, {Martinez-Rubi}, {Massart}, {Meynadier}, {Mignot}, {Munari}, {Nguyen}, {Nordlander}, {Ocvirk}, {O'Flaherty}, {Olias Sanz}, {Ortiz}, {Osorio}, {Oszkiewicz}, {Ouzounis}, {Palmer}, {Park}, {Pasquato}, {Peltzer}, {Peralta}, {P{\'e}turaud}, {Pieniluoma}, {Pigozzi}, {Poels}, {Prat}, {Prod'homme}, {Raison}, {Rebordao}, {Risquez}, {Rocca-Volmerange}, {Rosen},
  {Ruiz-Fuertes}, {Russo}, {Sembay}, {Serraller Vizcaino}, {Short}, {Siebert}, {Silva}, {Sinachopoulos}, {Slezak}, {Soffel}, {Sosnowska}, {Strai{\v{z}}ys}, {ter Linden}, {Terrell}, {Theil}, {Tiede}, {Troisi}, {Tsalmantza}, {Tur}, {Vaccari}, {Vachier}, {Valles}, {Van Hamme}, {Veltz}, {Virtanen}, {Wallut}, {Wichmann}, {Wilkinson}, {Ziaeepour}, \& {Zschocke}}]{gaia}
{Gaia Collaboration}, {Prusti}, T., {de Bruijne}, J.~H.~J., {et~al.} 2016, \aap, 595, A1, \dodoi{10.1051/0004-6361/201629272}

\bibitem[{{Gaia Collaboration} {et~al.}(2018){Gaia Collaboration}, {Babusiaux}, {van Leeuwen}, {Barstow}, {Jordi}, {Vallenari}, {Bossini}, {Bressan}, {Cantat-Gaudin}, {van Leeuwen}, {Brown}, {Prusti}, {de Bruijne}, {Bailer-Jones}, {Biermann}, {Evans}, {Eyer}, {Jansen}, {Klioner}, {Lammers}, {Lindegren}, {Luri}, {Mignard}, {Panem}, {Pourbaix}, {Randich}, {Sartoretti}, {Siddiqui}, {Soubiran}, {Walton}, {Arenou}, {Bastian}, {Cropper}, {Drimmel}, {Katz}, {Lattanzi}, {Bakker}, {Cacciari}, {Casta{\~n}eda}, {Chaoul}, {Cheek}, {De Angeli}, {Fabricius}, {Guerra}, {Holl}, {Masana}, {Messineo}, {Mowlavi}, {Nienartowicz}, {Panuzzo}, {Portell}, {Riello}, {Seabroke}, {Tanga}, {Th{\'e}venin}, {Gracia-Abril}, {Comoretto}, {Garcia-Reinaldos}, {Teyssier}, {Altmann}, {Andrae}, {Audard}, {Bellas-Velidis}, {Benson}, {Berthier}, {Blomme}, {Burgess}, {Busso}, {Carry}, {Cellino}, {Clementini}, {Clotet}, {Creevey}, {Davidson}, {De Ridder}, {Delchambre}, {Dell'Oro}, {Ducourant}, {Fern{\'a}ndez-Hern{\'a}ndez}, {Fouesneau},
  {Fr{\'e}mat}, {Galluccio}, {Garc{\'\i}a-Torres}, {Gonz{\'a}lez-N{\'u}{\~n}ez}, {Gonz{\'a}lez-Vidal}, {Gosset}, {Guy}, {Halbwachs}, {Hambly}, {Harrison}, {Hern{\'a}ndez}, {Hestroffer}, {Hodgkin}, {Hutton}, {Jasniewicz}, {Jean-Antoine-Piccolo}, {Jordan}, {Korn}, {Krone-Martins}, {Lanzafame}, {Lebzelter}, {L{\"o}ffler}, {Manteiga}, {Marrese}, {Mart{\'\i}n-Fleitas}, {Moitinho}, {Mora}, {Muinonen}, {Osinde}, {Pancino}, {Pauwels}, {Petit}, {Recio-Blanco}, {Richards}, {Rimoldini}, {Robin}, {Sarro}, {Siopis}, {Smith}, {Sozzetti}, {S{\"u}veges}, {Torra}, {van Reeven}, {Abbas}, {Abreu Aramburu}, {Accart}, {Aerts}, {Altavilla}, {{\'A}lvarez}, {Alvarez}, {Alves}, {Anderson}, {Andrei}, {Anglada Varela}, {Antiche}, {Antoja}, {Arcay}, {Astraatmadja}, {Bach}, {Baker}, {Balaguer-N{\'u}{\~n}ez}, {Balm}, {Barache}, {Barata}, {Barbato}, {Barblan}, {Barklem}, {Barrado}, {Barros}, {Bartholom{\'e} Mu{\~n}oz}, {Bassilana}, {Becciani}, {Bellazzini}, {Berihuete}, {Bertone}, {Bianchi}, {Bienaym{\'e}}, {Blanco-Cuaresma}, {Boch},
  {Boeche}, {Bombrun}, {Borrachero}, {Bouquillon}, {Bourda}, {Bragaglia}, {Bramante}, {Breddels}, {Brouillet}, {Br{\"u}semeister}, {Brugaletta}, {Bucciarelli}, {Burlacu}, {Busonero}, {Butkevich}, {Buzzi}, {Caffau}, {Cancelliere}, {Cannizzaro}, {Carballo}, {Carlucci}, {Carrasco}, {Casamiquela}, {Castellani}, {Castro-Ginard}, {Charlot}, {Chemin}, {Chiavassa}, {Cocozza}, {Costigan}, {Cowell}, {Crifo}, {Crosta}, {Crowley}, {Cuypers}, {Dafonte}, {Damerdji}, {Dapergolas}, {David}, {David}, {de Laverny}, {De Luise}, {De March}, {de Martino}, {de Souza}, {de Torres}, {Debosscher}, {del Pozo}, {Delbo}, {Delgado}, {Delgado}, {Diakite}, {Diener}, {Distefano}, {Dolding}, {Drazinos}, {Dur{\'a}n}, {Edvardsson}, {Enke}, {Eriksson}, {Esquej}, {Eynard Bontemps}, {Fabre}, {Fabrizio}, {Faigler}, {Falc{\~a}o}, {Farr{\`a}s Casas}, {Federici}, {Fedorets}, {Fernique}, {Figueras}, {Filippi}, {Findeisen}, {Fonti}, {Fraile}, {Fraser}, {Fr{\'e}zouls}, {Gai}, {Galleti}, {Garabato}, {Garc{\'\i}a-Sedano}, {Garofalo}, {Garralda}, {Gavel},
  {Gavras}, {Gerssen}, {Geyer}, {Giacobbe}, {Gilmore}, {Girona}, {Giuffrida}, {Glass}, {Gomes}, {Granvik}, {Gueguen}, {Guerrier}, {Guiraud}, {Guti{\'e}}, {Haigron}, {Hatzidimitriou}, {Hauser}, {Haywood}, {Heiter}, {Helmi}, {Heu}, {Hilger}, {Hobbs}, {Hofmann}, {Holland}, {Huckle}, {Hypki}, {Icardi}, {Jan{\ss}en}, {Jevardat de Fombelle}, {Jonker}, {Juh{\'a}sz}, {Julbe}, {Karampelas}, {Kewley}, {Klar}, {Kochoska}, {Kohley}, {Kolenberg}, {Kontizas}, {Kontizas}, {Koposov}, {Kordopatis}, {Kostrzewa-Rutkowska}, {Koubsky}, {Lambert}, {Lanza}, {Lasne}, {Lavigne}, {Le Fustec}, {Le Poncin-Lafitte}, {Lebreton}, {Leccia}, {Leclerc}, {Lecoeur-Taibi}, {Lenhardt}, {Leroux}, {Liao}, {Licata}, {Lindstr{\o}m}, {Lister}, {Livanou}, {Lobel}, {L{\'o}pez}, {Managau}, {Mann}, {Mantelet}, {Marchal}, {Marchant}, {Marconi}, {Marinoni}, {Marschalk{\'o}}, {Marshall}, {Martino}, {Marton}, {Mary}, {Massari}, {Matijevi{\v{c}}}, {Mazeh}, {McMillan}, {Messina}, {Michalik}, {Millar}, {Molina}, {Molinaro}, {Moln{\'a}r}, {Montegriffo}, {Mor},
  {Morbidelli}, {Morel}, {Morris}, {Mulone}, {Muraveva}, {Musella}, {Nelemans}, {Nicastro}, {Noval}, {O'Mullane}, {Ord{\'e}novic}, {Ord{\'o}{\~n}ez-Blanco}, {Osborne}, {Pagani}, {Pagano}, {Pailler}, {Palacin}, {Palaversa}, {Panahi}, {Pawlak}, {Piersimoni}, {Pineau}, {Plachy}, {Plum}, {Poggio}, {Poujoulet}, {Pr{\v{s}}a}, {Pulone}, {Racero}, {Ragaini}, {Rambaux}, {Ramos-Lerate}, {Regibo}, {Reyl{\'e}}, {Riclet}, {Ripepi}, {Riva}, {Rivard}, {Rixon}, {Roegiers}, {Roelens}, {Romero-G{\'o}mez}, {Rowell}, {Royer}, {Ruiz-Dern}, {Sadowski}, {Sagrist{\`a} Sell{\'e}s}, {Sahlmann}, {Salgado}, {Salguero}, {Sanna}, {Santana-Ros}, {Sarasso}, {Savietto}, {Schultheis}, {Sciacca}, {Segol}, {Segovia}, {S{\'e}gransan}, {Shih}, {Siltala}, {Silva}, {Smart}, {Smith}, {Solano}, {Solitro}, {Sordo}, {Soria Nieto}, {Souchay}, {Spagna}, {Spoto}, {Stampa}, {Steele}, {Steidelm{\"u}ller}, {Stephenson}, {Stoev}, {Suess}, {Surdej}, {Szabados}, {Szegedi-Elek}, {Tapiador}, {Taris}, {Tauran}, {Taylor}, {Teixeira}, {Terrett}, {Teyssandier},
  {Thuillot}, {Titarenko}, {Torra Clotet}, {Turon}, {Ulla}, {Utrilla}, {Uzzi}, {Vaillant}, {Valentini}, {Valette}, {van Elteren}, {Van Hemelryck}, {Vaschetto}, {Vecchiato}, {Veljanoski}, {Viala}, {Vicente}, {Vogt}, {von Essen}, {Voss}, {Votruba}, {Voutsinas}, {Walmsley}, {Weiler}, {Wertz}, {Wevers}, {Wyrzykowski}, {Yoldas}, {{\v{Z}}erjal}, {Ziaeepour}, {Zorec}, {Zschocke}, {Zucker}, {Zurbach}, \& {Zwitter}}]{gaia2018HR}
{Gaia Collaboration}, {Babusiaux}, C., {van Leeuwen}, F., {et~al.} 2018, \aap, 616, A10, \dodoi{10.1051/0004-6361/201832843}

\bibitem[{{Gaia Collaboration} {et~al.}(2021){Gaia Collaboration}, {Brown}, {Vallenari}, {Prusti}, {de Bruijne}, {Babusiaux}, {Biermann}, {Creevey}, {Evans}, {Eyer}, {Hutton}, {Jansen}, {Jordi}, {Klioner}, {Lammers}, {Lindegren}, {Luri}, {Mignard}, {Panem}, {Pourbaix}, {Randich}, {Sartoretti}, {Soubiran}, {Walton}, {Arenou}, {Bailer-Jones}, {Bastian}, {Cropper}, {Drimmel}, {Katz}, {Lattanzi}, {van Leeuwen}, {Bakker}, {Cacciari}, {Casta{\~n}eda}, {De Angeli}, {Ducourant}, {Fabricius}, {Fouesneau}, {Fr{\'e}mat}, {Guerra}, {Guerrier}, {Guiraud}, {Jean-Antoine Piccolo}, {Masana}, {Messineo}, {Mowlavi}, {Nicolas}, {Nienartowicz}, {Pailler}, {Panuzzo}, {Riclet}, {Roux}, {Seabroke}, {Sordo}, {Tanga}, {Th{\'e}venin}, {Gracia-Abril}, {Portell}, {Teyssier}, {Altmann}, {Andrae}, {Bellas-Velidis}, {Benson}, {Berthier}, {Blomme}, {Brugaletta}, {Burgess}, {Busso}, {Carry}, {Cellino}, {Cheek}, {Clementini}, {Damerdji}, {Davidson}, {Delchambre}, {Dell'Oro}, {Fern{\'a}ndez-Hern{\'a}ndez}, {Galluccio}, {Garc{\'\i}a-Lario},
  {Garcia-Reinaldos}, {Gonz{\'a}lez-N{\'u}{\~n}ez}, {Gosset}, {Haigron}, {Halbwachs}, {Hambly}, {Harrison}, {Hatzidimitriou}, {Heiter}, {Hern{\'a}ndez}, {Hestroffer}, {Hodgkin}, {Holl}, {Jan{\ss}en}, {Jevardat de Fombelle}, {Jordan}, {Krone-Martins}, {Lanzafame}, {L{\"o}ffler}, {Lorca}, {Manteiga}, {Marchal}, {Marrese}, {Moitinho}, {Mora}, {Muinonen}, {Osborne}, {Pancino}, {Pauwels}, {Petit}, {Recio-Blanco}, {Richards}, {Riello}, {Rimoldini}, {Robin}, {Roegiers}, {Rybizki}, {Sarro}, {Siopis}, {Smith}, {Sozzetti}, {Ulla}, {Utrilla}, {van Leeuwen}, {van Reeven}, {Abbas}, {Abreu Aramburu}, {Accart}, {Aerts}, {Aguado}, {Ajaj}, {Altavilla}, {{\'A}lvarez}, {{\'A}lvarez Cid-Fuentes}, {Alves}, {Anderson}, {Anglada Varela}, {Antoja}, {Audard}, {Baines}, {Baker}, {Balaguer-N{\'u}{\~n}ez}, {Balbinot}, {Balog}, {Barache}, {Barbato}, {Barros}, {Barstow}, {Bartolom{\'e}}, {Bassilana}, {Bauchet}, {Baudesson-Stella}, {Becciani}, {Bellazzini}, {Bernet}, {Bertone}, {Bianchi}, {Blanco-Cuaresma}, {Boch}, {Bombrun}, {Bossini},
  {Bouquillon}, {Bragaglia}, {Bramante}, {Breedt}, {Bressan}, {Brouillet}, {Bucciarelli}, {Burlacu}, {Busonero}, {Butkevich}, {Buzzi}, {Caffau}, {Cancelliere}, {C{\'a}novas}, {Cantat-Gaudin}, {Carballo}, {Carlucci}, {Carnerero}, {Carrasco}, {Casamiquela}, {Castellani}, {Castro-Ginard}, {Castro Sampol}, {Chaoul}, {Charlot}, {Chemin}, {Chiavassa}, {Cioni}, {Comoretto}, {Cooper}, {Cornez}, {Cowell}, {Crifo}, {Crosta}, {Crowley}, {Dafonte}, {Dapergolas}, {David}, {David}, {de Laverny}, {De Luise}, {De March}, {De Ridder}, {de Souza}, {de Teodoro}, {de Torres}, {del Peloso}, {del Pozo}, {Delbo}, {Delgado}, {Delgado}, {Delisle}, {Di Matteo}, {Diakite}, {Diener}, {Distefano}, {Dolding}, {Eappachen}, {Edvardsson}, {Enke}, {Esquej}, {Fabre}, {Fabrizio}, {Faigler}, {Fedorets}, {Fernique}, {Fienga}, {Figueras}, {Fouron}, {Fragkoudi}, {Fraile}, {Franke}, {Gai}, {Garabato}, {Garcia-Gutierrez}, {Garc{\'\i}a-Torres}, {Garofalo}, {Gavras}, {Gerlach}, {Geyer}, {Giacobbe}, {Gilmore}, {Girona}, {Giuffrida}, {Gomel}, {Gomez},
  {Gonzalez-Santamaria}, {Gonz{\'a}lez-Vidal}, {Granvik}, {Guti{\'e}rrez-S{\'a}nchez}, {Guy}, {Hauser}, {Haywood}, {Helmi}, {Hidalgo}, {Hilger}, {H{\l}adczuk}, {Hobbs}, {Holland}, {Huckle}, {Jasniewicz}, {Jonker}, {Juaristi Campillo}, {Julbe}, {Karbevska}, {Kervella}, {Khanna}, {Kochoska}, {Kontizas}, {Kordopatis}, {Korn}, {Kostrzewa-Rutkowska}, {Kruszy{\'n}ska}, {Lambert}, {Lanza}, {Lasne}, {Le Campion}, {Le Fustec}, {Lebreton}, {Lebzelter}, {Leccia}, {Leclerc}, {Lecoeur-Taibi}, {Liao}, {Licata}, {Lindstr{\o}m}, {Lister}, {Livanou}, {Lobel}, {Madrero Pardo}, {Managau}, {Mann}, {Marchant}, {Marconi}, {Marcos Santos}, {Marinoni}, {Marocco}, {Marshall}, {Martin Polo}, {Mart{\'\i}n-Fleitas}, {Masip}, {Massari}, {Mastrobuono-Battisti}, {Mazeh}, {McMillan}, {Messina}, {Michalik}, {Millar}, {Mints}, {Molina}, {Molinaro}, {Moln{\'a}r}, {Montegriffo}, {Mor}, {Morbidelli}, {Morel}, {Morris}, {Mulone}, {Munoz}, {Muraveva}, {Murphy}, {Musella}, {Noval}, {Ord{\'e}novic}, {Orr{\`u}}, {Osinde}, {Pagani}, {Pagano},
  {Palaversa}, {Palicio}, {Panahi}, {Pawlak}, {Pe{\~n}alosa Esteller}, {Penttil{\"a}}, {Piersimoni}, {Pineau}, {Plachy}, {Plum}, {Poggio}, {Poretti}, {Poujoulet}, {Pr{\v{s}}a}, {Pulone}, {Racero}, {Ragaini}, {Rainer}, {Raiteri}, {Rambaux}, {Ramos}, {Ramos-Lerate}, {Re Fiorentin}, {Regibo}, {Reyl{\'e}}, {Ripepi}, {Riva}, {Rixon}, {Robichon}, {Robin}, {Roelens}, {Rohrbasser}, {Romero-G{\'o}mez}, {Rowell}, {Royer}, {Rybicki}, {Sadowski}, {Sagrist{\`a} Sell{\'e}s}, {Sahlmann}, {Salgado}, {Salguero}, {Samaras}, {Sanchez Gimenez}, {Sanna}, {Santove{\~n}a}, {Sarasso}, {Schultheis}, {Sciacca}, {Segol}, {Segovia}, {S{\'e}gransan}, {Semeux}, {Shahaf}, {Siddiqui}, {Siebert}, {Siltala}, {Slezak}, {Smart}, {Solano}, {Solitro}, {Souami}, {Souchay}, {Spagna}, {Spoto}, {Steele}, {Steidelm{\"u}ller}, {Stephenson}, {S{\"u}veges}, {Szabados}, {Szegedi-Elek}, {Taris}, {Tauran}, {Taylor}, {Teixeira}, {Thuillot}, {Tonello}, {Torra}, {Torra}, {Turon}, {Unger}, {Vaillant}, {van Dillen}, {Vanel}, {Vecchiato}, {Viala}, {Vicente},
  {Voutsinas}, {Weiler}, {Wevers}, {Wyrzykowski}, {Yoldas}, {Yvard}, {Zhao}, {Zorec}, {Zucker}, {Zurbach}, \& {Zwitter}}]{gaiadr3}
{Gaia Collaboration}, {Brown}, A.~G.~A., {Vallenari}, A., {et~al.} 2021, \aap, 649, A1, \dodoi{10.1051/0004-6361/202039657}

\bibitem[{{Garofalo} {et~al.}(2022){Garofalo}, {Delgado}, {Sarro}, {Clementini}, {Muraveva}, {Marconi}, \& {Ripepi}}]{Garofalo2022}
{Garofalo}, A., {Delgado}, H.~E., {Sarro}, L.~M., {et~al.} 2022, \mnras, 513, 788, \dodoi{10.1093/mnras/stac735}

\bibitem[{{Green}(2018)}]{Green2018}
{Green}, G. 2018, The Journal of Open Source Software, 3, 695, \dodoi{10.21105/joss.00695}

\bibitem[{{Green} {et~al.}(2019){Green}, {Schlafly}, {Zucker}, {Speagle}, \& {Finkbeiner}}]{Green2019}
{Green}, G.~M., {Schlafly}, E., {Zucker}, C., {Speagle}, J.~S., \& {Finkbeiner}, D. 2019, \apj, 887, 93, \dodoi{10.3847/1538-4357/ab5362}

\bibitem[{{Gruberbauer} {et~al.}(2007){Gruberbauer}, {Kolenberg}, {Rowe}, {Huber}, {Matthews}, {Reegen}, {Kuschnig}, {Cameron}, {Kallinger}, {Weiss}, {Guenther}, {Moffat}, {Rucinski}, {Sasselov}, \& {Walker}}]{Gruberbauer2007}
{Gruberbauer}, M., {Kolenberg}, K., {Rowe}, J.~F., {et~al.} 2007, \mnras, 379, 1498, \dodoi{10.1111/j.1365-2966.2007.12042.x}

\bibitem[{Harris {et~al.}(2020)Harris, Millman, van~der Walt, Gommers, Virtanen, Cournapeau, Wieser, Taylor, Berg, Smith, Kern, Picus, Hoyer, van Kerkwijk, Brett, Haldane, del R{\'{i}}o, Wiebe, Peterson, G{\'{e}}rard-Marchant, Sheppard, Reddy, Weckesser, Abbasi, Gohlke, \& Oliphant}]{Numpy}
Harris, C.~R., Millman, K.~J., van~der Walt, S.~J., {et~al.} 2020, Nature, 585, 357, \dodoi{10.1038/s41586-020-2649-2}

\bibitem[{{Hattori} {et~al.}(2022){Hattori}, {Foreman-Mackey}, {Hogg}, {Montet}, {Angus}, {Pritchard}, {Curtis}, \& {Sch{\"o}lkopf}}]{Hattori2022}
{Hattori}, S., {Foreman-Mackey}, D., {Hogg}, D.~W., {et~al.} 2022, \aj, 163, 284, \dodoi{10.3847/1538-3881/ac625a}

\bibitem[{{Huang} \& {Koposov}(2022)}]{Huang2022}
{Huang}, K.-W., \& {Koposov}, S.~E. 2022, \mnras, 510, 3575, \dodoi{10.1093/mnras/stab3654}

\bibitem[{{Hunt} {et~al.}(2022){Hunt}, {Price-Whelan}, {Johnston}, \& {Darragh-Ford}}]{Hunt:2022}
{Hunt}, J. A.~S., {Price-Whelan}, A.~M., {Johnston}, K.~V., \& {Darragh-Ford}, E. 2022, \mnras, 516, L7, \dodoi{10.1093/mnrasl/slac082}

\bibitem[{Hunter(2007)}]{matplotlib}
Hunter, J.~D. 2007, Computing in Science \& Engineering, 9, 90, \dodoi{10.1109/MCSE.2007.55}

\bibitem[{{Iorio} \& {Belokurov}(2021)}]{Iorio2021}
{Iorio}, G., \& {Belokurov}, V. 2021, \mnras, 502, 5686, \dodoi{10.1093/mnras/stab005}

\bibitem[{{Jayasinghe} {et~al.}(2019){Jayasinghe}, {Stanek}, {Kochanek}, {Shappee}, {Holoien}, {Thompson}, {Prieto}, {Dong}, {Pawlak}, {Pejcha}, {Shields}, {Pojmanski}, {Otero}, {Britt}, \& {Will}}]{Jayasinghe2019}
{Jayasinghe}, T., {Stanek}, K.~Z., {Kochanek}, C.~S., {et~al.} 2019, \mnras, 486, 1907, \dodoi{10.1093/mnras/stz844}

\bibitem[{{Kochanek} {et~al.}(2017){Kochanek}, {Shappee}, {Stanek}, {Holoien}, {Thompson}, {Prieto}, {Dong}, {Shields}, {Will}, {Britt}, {Perzanowski}, \& {Pojma{\'n}ski}}]{asassn}
{Kochanek}, C.~S., {Shappee}, B.~J., {Stanek}, K.~Z., {et~al.} 2017, \pasp, 129, 104502, \dodoi{10.1088/1538-3873/aa80d9}

\bibitem[{{Lightkurve Collaboration} {et~al.}(2018){Lightkurve Collaboration}, {Cardoso}, {Hedges}, {Gully-Santiago}, {Saunders}, {Cody}, {Barclay}, {Hall}, {Sagear}, {Turtelboom}, {Zhang}, {Tzanidakis}, {Mighell}, {Coughlin}, {Bell}, {Berta-Thompson}, {Williams}, {Dotson}, \& {Barentsen}}]{lightkurve2018}
{Lightkurve Collaboration}, {Cardoso}, J.~V.~d.~M., {Hedges}, C., {et~al.} 2018, {Lightkurve: Kepler and TESS time series analysis in Python}, Astrophysics Source Code Library.
\newblock \doeprint{1812.013}

\bibitem[{{Marconi} {et~al.}(2015){Marconi}, {Coppola}, {Bono}, {Braga}, {Pietrinferni}, {Buonanno}, {Castellani}, {Musella}, {Ripepi}, \& {Stellingwerf}}]{Marconi2015}
{Marconi}, M., {Coppola}, G., {Bono}, G., {et~al.} 2015, \apj, 808, 50, \dodoi{10.1088/0004-637X/808/1/50}

\bibitem[{Mateu(2024)}]{Mateu2024}
Mateu, C. 2024, Research Notes of the AAS, 8, 85, \dodoi{10.3847/2515-5172/ad3540}

\bibitem[{{Mateu} {et~al.}(2020){Mateu}, {Holl}, {De Ridder}, \& {Rimoldini}}]{Mateu2020}
{Mateu}, C., {Holl}, B., {De Ridder}, J., \& {Rimoldini}, L. 2020, \mnras, 496, 3291, \dodoi{10.1093/mnras/staa1676}

\bibitem[{McKinney {et~al.}(2010)}]{pandas}
McKinney, W., {et~al.} 2010, in Proceedings of the 9th Python in Science Conference, Vol. 445, Austin, TX, 51--56

\bibitem[{{Moln{\'a}r} {et~al.}(2012){Moln{\'a}r}, {Koll{\'a}th}, {Szab{\'o}}, {Bryson}, {Kolenberg}, {Mullally}, \& {Thompson}}]{Molnar2012}
{Moln{\'a}r}, L., {Koll{\'a}th}, Z., {Szab{\'o}}, R., {et~al.} 2012, \apjl, 757, L13, \dodoi{10.1088/2041-8205/757/1/L13}

\bibitem[{{Moln{\'a}r} {et~al.}(2022){Moln{\'a}r}, {B{\'o}di}, {P{\'a}l}, {Bhardwaj}, {Hambsch}, {Benk{\H{o}}}, {Derekas}, {Ebadi}, {Joyce}, {Hasanzadeh}, {Kolenberg}, {Lund}, {Nemec}, {Netzel}, {Ngeow}, {Pepper}, {Plachy}, {Prudil}, {Siverd}, {Skarka}, {Smolec}, {S{\'o}dor}, {Sylla}, {Szab{\'o}}, {Szab{\'o}}, {Kjeldsen}, {Christensen-Dalsgaard}, \& {Ricker}}]{Molnar2022}
{Moln{\'a}r}, L., {B{\'o}di}, A., {P{\'a}l}, A., {et~al.} 2022, \apjs, 258, 8, \dodoi{10.3847/1538-4365/ac2ee2}

\bibitem[{{Monnier} {et~al.}(2007){Monnier}, {Zhao}, {Pedretti}, {Thureau}, {Ireland}, {Muirhead}, {Berger}, {Millan-Gabet}, {Van Belle}, {ten Brummelaar}, {McAlister}, {Ridgway}, {Turner}, {Sturmann}, {Sturmann}, \& {Berger}}]{Monnier2007}
{Monnier}, J.~D., {Zhao}, M., {Pedretti}, E., {et~al.} 2007, Science, 317, 342, \dodoi{10.1126/science.1143205}

\bibitem[{{Murphy} {et~al.}(2015){Murphy}, {Bedding}, {Niemczura}, {Kurtz}, \& {Smalley}}]{Murphy2015}
{Murphy}, S.~J., {Bedding}, T.~R., {Niemczura}, E., {Kurtz}, D.~W., \& {Smalley}, B. 2015, \mnras, 447, 3948, \dodoi{10.1093/mnras/stu2749}

\bibitem[{{Murphy} {et~al.}(2019){Murphy}, {Hey}, {Van Reeth}, \& {Bedding}}]{Murphy2019}
{Murphy}, S.~J., {Hey}, D., {Van Reeth}, T., \& {Bedding}, T.~R. 2019, \mnras, 485, 2380, \dodoi{10.1093/mnras/stz590}

\bibitem[{{Narloch} {et~al.}(2019){Narloch}, {Pietrzy{\'n}ski}, {Ko{\l}aczkowski}, {Smolec}, {G{\'o}rski}, {Kubiak}, {Udalski}, {Soszy{\'n}ski}, {Graczyk}, {Gieren}, {Karczmarek}, {Zgirski}, {Wielg{\'o}rski}, {Suchomska}, {Pilecki}, {Taormina}, \& {Ka{\l}uszy{\'n}ski}}]{Narloch2019}
{Narloch}, W., {Pietrzy{\'n}ski}, G., {Ko{\l}aczkowski}, Z., {et~al.} 2019, \mnras, 489, 3285, \dodoi{10.1093/mnras/stz2112}

\bibitem[{{Netzel} {et~al.}(2023){Netzel}, {Moln{\'a}r}, {Plachy}, \& {Benk{\H{o}}}}]{Netzel2023}
{Netzel}, H., {Moln{\'a}r}, L., {Plachy}, E., \& {Benk{\H{o}}}, J.~M. 2023, \aap, 677, A177, \dodoi{10.1051/0004-6361/202245634}

\bibitem[{{Ochsenbein} {et~al.}(2000){Ochsenbein}, {Bauer}, \& {Marcout}}]{vizier}
{Ochsenbein}, F., {Bauer}, P., \& {Marcout}, J. 2000, \aaps, 143, 23, \dodoi{10.1051/aas:2000169}

\bibitem[{{Preston}(1959)}]{Preston1959}
{Preston}, G.~W. 1959, \apj, 130, 507, \dodoi{10.1086/146743}

\bibitem[{{Price-Whelan}(2017)}]{gala2017}
{Price-Whelan}, A.~M. 2017, The Journal of Open Source Software, 2, 388, \dodoi{10.21105/joss.00388}

\bibitem[{{Price-Whelan} {et~al.}(2018){Price-Whelan}, {Sip{\H{o}}cz}, {G{\"u}nther}, {Lim}, {Crawford}, {Conseil}, {Shupe}, {Craig}, {Dencheva}, {Ginsburg}, {VanderPlas}, {Bradley}, {P{\'e}rez-Su{\'a}rez}, {de Val-Borro}, {Paper Contributors}, {Aldcroft}, {Cruz}, {Robitaille}, {Tollerud}, {Coordination Committee}, {Ardelean}, {Babej}, {Bach}, {Bachetti}, {Bakanov}, {Bamford}, {Barentsen}, {Barmby}, {Baumbach}, {Berry}, {Biscani}, {Boquien}, {Bostroem}, {Bouma}, {Brammer}, {Bray}, {Breytenbach}, {Buddelmeijer}, {Burke}, {Calderone}, {Cano Rodr{\'\i}guez}, {Cara}, {Cardoso}, {Cheedella}, {Copin}, {Corrales}, {Crichton}, {D{\textquoteright}Avella}, {Deil}, {Depagne}, {Dietrich}, {Donath}, {Droettboom}, {Earl}, {Erben}, {Fabbro}, {Ferreira}, {Finethy}, {Fox}, {Garrison}, {Gibbons}, {Goldstein}, {Gommers}, {Greco}, {Greenfield}, {Groener}, {Grollier}, {Hagen}, {Hirst}, {Homeier}, {Horton}, {Hosseinzadeh}, {Hu}, {Hunkeler}, {Ivezi{\'c}}, {Jain}, {Jenness}, {Kanarek}, {Kendrew}, {Kern}, {Kerzendorf}, {Khvalko},
  {King}, {Kirkby}, {Kulkarni}, {Kumar}, {Lee}, {Lenz}, {Littlefair}, {Ma}, {Macleod}, {Mastropietro}, {McCully}, {Montagnac}, {Morris}, {Mueller}, {Mumford}, {Muna}, {Murphy}, {Nelson}, {Nguyen}, {Ninan}, {N{\"o}the}, {Ogaz}, {Oh}, {Parejko}, {Parley}, {Pascual}, {Patil}, {Patil}, {Plunkett}, {Prochaska}, {Rastogi}, {Reddy Janga}, {Sabater}, {Sakurikar}, {Seifert}, {Sherbert}, {Sherwood-Taylor}, {Shih}, {Sick}, {Silbiger}, {Singanamalla}, {Singer}, {Sladen}, {Sooley}, {Sornarajah}, {Streicher}, {Teuben}, {Thomas}, {Tremblay}, {Turner}, {Terr{\'o}n}, {van Kerkwijk}, {de la Vega}, {Watkins}, {Weaver}, {Whitmore}, {Woillez}, {Zabalza}, \& {Contributors}}]{astropy:2018}
{Price-Whelan}, A.~M., {Sip{\H{o}}cz}, B.~M., {G{\"u}nther}, H.~M., {et~al.} 2018, \aj, 156, 123, \dodoi{10.3847/1538-3881/aabc4f}

\bibitem[{{Prudil} {et~al.}(2020){Prudil}, {D{\'e}k{\'a}ny}, {Grebel}, \& {Kunder}}]{Prudil2020}
{Prudil}, Z., {D{\'e}k{\'a}ny}, I., {Grebel}, E.~K., \& {Kunder}, A. 2020, \mnras, 492, 3408, \dodoi{10.1093/mnras/staa046}

\bibitem[{{Reese} {et~al.}(2017){Reese}, {Ligni{\`e}res}, {Ballot}, {Dupret}, {Barban}, {van't Veer-Menneret}, \& {MacGregor}}]{Reese2017}
{Reese}, D.~R., {Ligni{\`e}res}, F., {Ballot}, J., {et~al.} 2017, \aap, 601, A130, \dodoi{10.1051/0004-6361/201321264}

\bibitem[{{Ricker} {et~al.}(2015){Ricker}, {Winn}, {Vanderspek}, {Latham}, {Bakos}, {Bean}, {Berta-Thompson}, {Brown}, {Buchhave}, {Butler}, {Butler}, {Chaplin}, {Charbonneau}, {Christensen-Dalsgaard}, {Clampin}, {Deming}, {Doty}, {De Lee}, {Dressing}, {Dunham}, {Endl}, {Fressin}, {Ge}, {Henning}, {Holman}, {Howard}, {Ida}, {Jenkins}, {Jernigan}, {Johnson}, {Kaltenegger}, {Kawai}, {Kjeldsen}, {Laughlin}, {Levine}, {Lin}, {Lissauer}, {MacQueen}, {Marcy}, {McCullough}, {Morton}, {Narita}, {Paegert}, {Palle}, {Pepe}, {Pepper}, {Quirrenbach}, {Rinehart}, {Sasselov}, {Sato}, {Seager}, {Sozzetti}, {Stassun}, {Sullivan}, {Szentgyorgyi}, {Torres}, {Udry}, \& {Villasenor}}]{TESS}
{Ricker}, G.~R., {Winn}, J.~N., {Vanderspek}, R., {et~al.} 2015, Journal of Astronomical Telescopes, Instruments, and Systems, 1, 014003, \dodoi{10.1117/1.JATIS.1.1.014003}

\bibitem[{{Saio} {et~al.}(2018){Saio}, {Bedding}, {Kurtz}, {Murphy}, {Antoci}, {Shibahashi}, {Li}, \& {Takata}}]{Saio2018}
{Saio}, H., {Bedding}, T.~R., {Kurtz}, D.~W., {et~al.} 2018, \mnras, 477, 2183, \dodoi{10.1093/mnras/sty784}

\bibitem[{{Sanders}(2012)}]{Sanders2012}
{Sanders}, J. 2012, \mnras, 426, 128, \dodoi{10.1111/j.1365-2966.2012.21698.x}

\bibitem[{{Sesar} {et~al.}(2017){Sesar}, {Hernitschek}, {Mitrovi{\'c}}, {Ivezi{\'c}}, {Rix}, {Cohen}, {Bernard}, {Grebel}, {Martin}, {Schlafly}, {Burgett}, {Draper}, {Flewelling}, {Kaiser}, {Kudritzki}, {Magnier}, {Metcalfe}, {Tonry}, \& {Waters}}]{Sesar2017_ps1_rrl}
{Sesar}, B., {Hernitschek}, N., {Mitrovi{\'c}}, S., {et~al.} 2017, \aj, 153, 204, \dodoi{10.3847/1538-3881/aa661b}

\bibitem[{{Smith}(1995)}]{Smith1995}
{Smith}, H.~A. 1995, Cambridge Astrophysics Series, 27

\bibitem[{{Stellingwerf}(1982)}]{Stellingwerf1982}
{Stellingwerf}, R.~F. 1982, \apj, 262, 330, \dodoi{10.1086/160425}

\bibitem[{{Wallace} {et~al.}(2019){Wallace}, {Hartman}, {Bakos}, \& {Bhatti}}]{Wallace2019}
{Wallace}, J.~J., {Hartman}, J.~D., {Bakos}, G.~{\'A}., \& {Bhatti}, W. 2019, \apjl, 870, L7, \dodoi{10.3847/2041-8213/aaf8ac}

\bibitem[{{Wenger} {et~al.}(2000){Wenger}, {Ochsenbein}, {Egret}, {Dubois}, {Bonnarel}, {Borde}, {Genova}, {Jasniewicz}, {Lalo{\"e}}, {Lesteven}, \& {Monier}}]{simbad}
{Wenger}, M., {Ochsenbein}, F., {Egret}, D., {et~al.} 2000, \aaps, 143, 9, \dodoi{10.1051/aas:2000332}

\bibitem[{{Zinn} {et~al.}(2020){Zinn}, {Chen}, {Layden}, \& {Casetti-Dinescu}}]{Zinn2020}
{Zinn}, R., {Chen}, X., {Layden}, A.~C., \& {Casetti-Dinescu}, D.~I. 2020, \mnras, 492, 2161, \dodoi{10.1093/mnras/stz3580}

\end{thebibliography}
\bibliographystyle{aasjournal}



\end{document}